\documentclass[a4paper,12pt]{article}
\usepackage{latexsym}
\usepackage{amssymb}
\usepackage{amsfonts} 
\usepackage{amsmath}
\usepackage[mathscr]{euscript}

\title{\textsf{Lowest energy states in nonrelativistic QED: atoms and
ions in motion}}
\date{\empty}

\author{
Michael Loss\thanks{ Work partially supported by NSF grant DMS 03-00349}
,
Tadahiro Miyao\thanks{This work was supported  by Japan Society for the Promotion of
Science (JSPS).},
  and  Herbert Spohn\footnotemark[3]\\
\footnotemark[1] {\it School of Mathematics, IAS, Princeton NJ, 08540}\\
 {\it Permanent adress: School of Mathematics,}\\{\it  Georgia Institute of Technology}\\
{\it  Atlanta, GA 30332, USA}\\
\footnotemark[2] \footnotemark[3] {\it Zentrum Mathematik,}
{\it Technische Universit\"at M\"unchen,}\\ 
{\it  D-85747 Garching, Germany}\\
e-mail: \footnotemark[1] \texttt{loss@math.gatech.edu,}\\
\texttt{
\footnotemark[2] miyao@ma.tum.de,
\footnotemark[3] spohn@ma.tum.de } 
}

\newcommand{\one}{{\mathchoice {\rm 1\mskip-4mu l} {\rm 1\mskip-4mu l}
{\rm 1\mskip-4.5mu l} {\rm 1\mskip-5mu l}}}
\newcommand{\h}{\mathfrak{h}}
\newcommand{\Hil}{\mathcal{H}}
\newcommand{\ex}{\mathrm{e}}
\newcommand{\D}{\mathrm{dom}}

\newcommand{\im}{\mathrm{i}}
\newcommand{\Fock}{\mathfrak{F}}
\newcommand{\Ffin}{\mathfrak{F}_{\mathrm{fin}}}

\newcommand{\dG}{\mathrm{d}\Gamma}
\newcommand{\Num}{N_{\mathrm{f}}}

\newcommand{\hotimes}{\widehat{\otimes}}

\newcommand{\la}{\langle}
\newcommand{\ra}{\rangle}

\newcommand{\wlim}{\mbox{$\mathrm{w}$-$\displaystyle\lim_{n\to\infty}$}}
\newcommand{\slim}{\mbox{$\mathrm{s}$-$\displaystyle\lim_{n\to\infty}$}}

\newcommand{\BbbR}{\mathbb{R}}
\newcommand{\BbbN}{\mathbb{N}}

\newcommand{\BbbC}{\mathbb{C}}
\newcommand{\vepsilon}{\varepsilon}
\newcommand{\vphi}{\varphi}
\newcommand{\Pt}{P_{\mathrm{tot}}}
\newcommand{\Pf}{P_{\mathrm{f}}}
\newcommand{\Hf}{H_{\mathrm{f}}}
\newcommand{\Mass}{m}
\newcommand{\Gr}{\Psi_{P,\Mass}}
\newcommand{\Res}{\mathcal{R}_{P,\Mass}(k)}
\newcommand{\nmass}{m_0}
\newcommand{\Nm}{m_{\mathrm{n}}}
\newcommand{\emass}{m_{\mathrm{e}}}
\newcommand{\tmass}{m_{\mathrm{tot}}}
\newcommand{\Hilfin}{\mathcal{H}_{\mathrm{fin}}}
\newcommand{\mm}{\mathsf{m}}

\begin{document}

\newtheorem{define}{Definition}[section]
\newtheorem{Thm}[define]{Theorem}
\newtheorem{Prop}[define]{Proposition}
\newtheorem{lemm}[define]{Lemma}
\newtheorem{rem}[define]{Remark}
\newtheorem{assum}{Condition}
\newtheorem{example}{Example}
\newtheorem{coro}[define]{Corollary}

\maketitle

\begin{abstract}
Within the framework of nonrelativisitic quantum electrodynamics we consider a single
 nucleus and $N$ electrons  coupled to the radiation field. Since the total
 momentum $P$ is conserved, the Hamiltonian $H$ admits a fiber
 decomposition with respect to $P$ with fiber Hamiltonian  $H(P)$. A stable atom,
 resp. ion, means that the fiber Hamiltonian $H(P)$ has an
 eigenvalue at the bottom of its spectrum. We establish  the existence of
 a  ground state for  $H(P)$ under (i) an  explicit bound on $P$, 
(ii) a binding condition, and (iii) an energy inequality. The
 binding  condition is proven to hold for a  heavy nucleus and the
 energy  inequality  for spinless electrons. 
\end{abstract}

\section{Introduction}
An atom, resp. ion,  consists  of a nucleus with  mass $m_{\mathrm{n}}$
and charge $Ze$  and $N$
electrons with  mass $\emass$ and charge $-e$. Within Schr\"odinger quantum
mechanics the atom is described by the Hamiltonian
\begin{align}
 h_N=-\frac{1}{2m_{\mathrm{n}}}\Delta_0-\sum_{j=1}^N\frac{1}{2\emass}\Delta_j+\sum_{1\le
 i< j\le
 N}\frac{e^2}{4\pi|x_i-x_j|}-\sum_{j=1}^N\frac{Ze^2}{4\pi|x_j-x_0|} ,\label{PHamiltonian}
\end{align}
where the units are such that $\hbar =1$. Here $x_0\in\BbbR^3$ is the
position of the nucleus, $x_j\in\BbbR^3$ the one of the $j$-th
electron,  $\Delta_j,\ j=0,\dots,N$, the corresponding Laplacian, and $m_{\mathrm{n}},m_{\mathrm{e}},Z>0$.
$h_N$ is regarded as an operator in $L^2(\BbbR^{3(N+1)})$. For the moment
we ignore the electron spin and Fermi statistics. $h_N$ commutes with the total momentum 
\begin{align}
 \Pt=\sum_{j=0}^Np_j,\ \ p_j=-\im \nabla_j.\label{PPtot}
\end{align}
Hence, trivially, $h_N$ has purely continuous spectrum. To investigate the
stability of the atom one  has to first transform to  atomic
 coordinates, see \cite{CFKS}. Then $h_N$ is  written as the direct integral
\begin{align}
 h_N=\int_{\BbbR^3}^{\oplus}h(P)\, \mathrm{d}P.\label{PDirect}
\end{align}
$h(P)$ is the Hamiltonian at fixed total momentum $P$ and has the form
\begin{align}
h(P)=\frac{1}{2\tmass}P^2+\tilde{h}
\end{align}
with $\tmass=m_{\mathrm{n}}+N\emass$. $\tilde{h}$ is independent of $P$
and  acts on
$L^2(\BbbR^{3N})$,    its precise form can  be found  in Equation (\ref{CenterRM}).
 The stability  of an atom    is thus reduced to prove that 
$\tilde{h}$ has an eigenvalue at the  bottom of its spectrum. By a
famous result of Zhislin \cite{HunSig}, see  also \cite{Zhislin}  such a property holds provided
$N<Z+1$. The existence of negatively charged ions is a much more tricky
business. We refer to \cite{ARV} for a survey.
Note  that a stable atom can move at any speed, since the center of
mass kinetic energy is proportional to $P^2$.

   The Coulomb interaction between the charges results from the coupling to
the Maxwell field and,  in a full quantum theory, also the electromagnetic
field has to be quantized. While ultimately such a path leads to
relativistic QED, for the present paper we settle at the
nonrelativistic version, which has only electrons, nuclei, and
photons as elementary objects.  Our  task is to understand, within the 
framework of nonrelativistic QED,
the stability of atoms and ions {\it in motion}.

We have to add to (\ref{PHamiltonian}) the field degrees of freedom and
the coupling of the charged particles to the field. For the present
study we consider a single nucleus with spin $0$ and $N$ spin $1/2$
electrons respecting  Fermi statistics, which results in the Hamiltonian
\begin{align}
  H=&\frac{1}{2\Nm}\big(-\im \nabla_0-Ze
 A(x_0)\big)^2+\sum_{j=1}^N\frac{1}{2\emass}\big\{\sigma_j\cdot\big(-\im\nabla_j+e
 A(x_j)\big)\big\}^2\nonumber\\
&+\sum_{1\le i<
  j\le
 N}\frac{e^2}{4\pi|x_i-x_j|}-\sum_{j=1}^N\frac{Ze^2}{4\pi|x_j-x_0|}+H_{\mathrm{f}}.
\label{QFullHamiltonian}
\end{align}
Here $\nabla_j$ is the gradient w.r.t. $x_j$ and $\sigma_j$ are the
 Pauli spin matrices of the $j$-th electron. 
 $A(x)$ is the quantized transverse vector potential and
$H_{\mathrm{f}}$ the energy of the photons with dispersion relation
$\omega(k)=|k|$, see (\ref{QVector}), (\ref{FreeH})  for a precise definition. 
We use units such that the speed of light $c=1$.  An
 ultraviolet  cutoff is always imposed. Otherwise $H$ would not be
properly defined. The infrared cutoff will be  studied  in detail.

As in the Schr\"odinger case, $H$ commutes with the total momentum
\begin{align}
\Pt=\Pf+\sum_{j=0}^N\big(-\im \nabla_j\big)
\end{align}
with  $\Pf$  the momentum of all photons.
Hence, if $H(P)$ denotes the Hamiltonian at fixed total momentum $P$, as
before one has the decomposition
\begin{align}
 H=\int^{\oplus}_{\BbbR^3}H(P)\, \mathrm{d}P. \label{QHamiltonian}
\end{align}
The problem  is to understand   under which  conditions $H(P)$ has an
eigenvalue at the bottom of its spectrum. Physically the corresponding
eigenstate describes a stable atom dressed with a photon cloud and in
motion with momentum $P$.

The case of a single charge, $N=0$ in our notation, has been studied  by
J. Fr\"ohlich in his ground-breaking thesis \cite{Froehlich}.
We borrow many of his insights.
For a more current study of the low energy regime  we refer to \cite{Chen}. Very recently, the case  of dressed atoms and ions, as governed by
Hamiltonian (\ref{QFullHamiltonian}), has been taken up by
 Amour, Grebert, and Guillot \cite{AGG}.
For $N\le Z$ they succeed to  prove that
$H(P)$ has a
ground state provided $|e|, |P|,$ and the ultraviolet cutoff are
sufficiently small.
Our aim here is to completely avoid such smallness assumptions,
by developing a   strategy  along  the lines
of \cite{LLG}.
There the author consider the hamiltonian $H_W=H+\sum_j W(x_j)$,
i.e. they add a confining one-body potenital $W$, and prove the
existence of absolute ground states provided a binding condition is
satisfied. $H_W$ does not conserve  the total momentum and a
decompostion as in (\ref{QHamiltonian}) is not possible. If the ground state of
$H_W$ exists, then the atom is at rest. Thus, in some sense, the results
in  \cite{LLG} cover the case when the total momentum vanishes.

The existence of a ground state for $H(P)$ will be established  under
four general assumptions. While their precise  form will stated in due course,
it should be helpful for the reader to understand their meaning in simple
terms, first. \medskip\\
{\it {\rm(i)} $|P|<P_{\mathrm{c}}$ (Cherenkov radiation)}. If a single charge
is accelerated to  a speed above the speed of light it emits Cherenkov
radiation and thereby slows down. Of course, physically, the electron has
to move in a medium where light propagates  with a speed less than $c$.
Our point is only that the model Hamiltonian (\ref{QFullHamiltonian}) knows about
Cherenkov radiation.
Mathematically Cherenkov radiation is reflected by the fact that
there exists some $P_{\mathrm{c}}$ such that $H(P)$ has a ground state
for $|P|<P_{\mathrm{c}}$, while $H(P)$ has no ground state for $|P|>P_{\mathrm{c}}$.
It has been established  already in \cite{Froehlich} that
$P_{\mathrm{c}}>(\sqrt{3}-1)\Nm$
for $N=0$, see also \cite{Spohn2}. Even for $N=0$, the converse statement, namely no ground state
for $P$ sufficiently large, is left as an open problem. To our
knowledge, the  only result
in this direction is provided in \cite{AMZ}, where the case  $N=0$ is
studied
 for   small coupling to a  scalar field.\medskip\\
{\it {\rm (ii)} Energy inequality}. Let $E(P)$ be  the bottom of the
spectrum of  $H(P)$. In our proof we need that
\begin{align}
E(0)\le E(P).\label{IntIneq}
\end{align}
Physically such a property appears to be obvious. 
But even for a single charge with spin we have no method to establish
(\ref{IntIneq}). We are equally at loss to include Fermi statistics. On
the other hand, in Section \ref{NoSS} we prove the inequality (\ref{IntIneq})
 for an arbitrary number of spinless charges satisfying Bose/Boltzmann
statistics.\medskip\\
{\it {\rm (iii)} Strictly positive binding energy.} 
Roughly speaking the binding condition states that energywise it is
more favorable to assemble all electrons near the nucleus compared to
having  one or several electrons placed at infinity. The presence of the
quantized radiation field complicates matter, but we will state a
suitable binding condition which reduces to the known condition when
the coupling to the field is ignored. Of course, to ensure the existence of a
ground state then requires to establish the binding condition.
We will prove it  for a heavy nucleus and, in greater generality,
for electrons without spin.
\medskip\\
{\it {\rm (iv)} Charge neutrality.} In $H$ of (\ref{QFullHamiltonian}) the
charge $e$ appears in the Coulomb potential and in the coupling to the
quantized transverse vector potential $A(x)$. After all, both originate
from the coupling to the Maxwell field. The particular splitting in 
(\ref{QFullHamiltonian}) is due
 to quantizing in the Coulomb gauge. Mathematically it is often
convenient to disregard such a link and to replace the Coulomb potential
by a general pair potential.  By neutrality we refer here  to the charges
entering in   the coupling to the vector potential.

If $Z=N$, then the quantized radiation field sees a neutral charge.
Thus, even for an atom in motion, the induced vector potential decays
faster
than $1/|x|$, which can indeed  be accomodated in Fock space. If $Z\neq N$, then
the quantized radiation field sees a non-zero charge.
If the atom is at rest, $P=0$,  classically the transverse vector field
vanishes and quantum mechanically $A(x)$  averaged in the ground state has a fast decay. On the other
hand if, $P\neq 0$,
then $A(x)$ decays as $1/|x|$, which cannot be accomodated in Fock space.
The putative physical ground state has an infinite number of (virtual)
photons. Therefore for $N\neq Z$  a ground state in Fock space can exist
only at $P=0$. Already for a single  charge, such a property is a rather delicate
phenomenon, see \cite{Chen} for  the best results available.
The results in \cite{AGG} and in our work are in agreement with such general reasoning.
For a neutral assembly of charges no infrared cutoff is needed. However
for a nonvanishing total charge we have  to impose a suitable infrared cutoff.
 \medskip\\
\ \ \  Perhaps more than in other papers, one of our difficulties concerns
the generality in which results are written out.
As guiding principle we adopt that at least one physically accepted
Hamiltonian
should be covered. This requires to work in space dimension $d=3$ and to
have  electrons with spin $1/2$. On the other hand the core of the
mathematical argument may  become hidden through over-explicit notation.
For example, we will replace the Coulomb potential by a general pair
potential from a class which  includes the Coulomb potential, of course.
The case of several spinless nuclei could be handled. If no statistics
is included, our proof carries over without changes. To include Bose
statistics requires extra efforts.

We provide a short outline of our paper.
 In Section 2 we define the
Hamiltonian for charges coupled to the Maxwell field
 and state the main result,  namely the 
  existence of a  ground state for  $H(P)$ for $P$ within  a suitable
range and under a strictly positive binding energy.
In case of an atom  with a heavy nucleus  
we provide explicit  bounds on the range of $P$ and
 for the validity of the binding condition.

The self-adjointness of  $H(P)$ for arbitrary
couplings and cutoffs is proven in Section 3.
As an essential input we use the same property for the full Hamiltonian
as established in \cite{Hiroshima1} by the use of  functional integral
techniques.

For a single charge some general properties of $E(P)=\inf
\mathrm{spec}(H(P))$ are demonstrated in \cite{Froehlich}.
In Section 4 we show how to extend them to an arbitrary number of
charges, in fact in a slightly strengthened version by  means of  a
variational technique.

In Section 5 we consider a non-zero photon mass by  replacing  in
$H_{\mathrm{f}}$ the dispersion relation $\omega(k)=|k|$ by
$\omega_m(k)=(k^2+m^2)^{1/2},\ m>0$. We assume  a strictly positive 
binding  energy and combine the methods in \cite{LLG} with the general
properties of $E(P)$ from Section 4. This yields the existence  of a
ground state for a suitable range of $P$'s. The remaining task is to
remove the infrared cutoff, i.e., $m\to 0$, see Section 6. For a neutral
system of charges the form factor $\hat{\vphi}(k)$ is allowed to have 
$\hat{\vphi}(0)=(2\pi)^{-3/2}$. For a non-neutral system the form factor
has to vanish as $\hat{\vphi}(k)\simeq |k|$ for small $k$.
Our  method is based on pull-through which yields a bound on the number
of soft photons and bounds on the derivative of the ground state wave
function with respect to the momenta of the photons.
 In the appendices we  collect some  technical
results.\medskip\\
{\sf Acknowledgements.}
T. Miyao  thanks  A. Arai, M. Griesemer, and  M. Hirokawa for useful
comments.
The present study was initiated  when M. Loss  visited the Zentrum Mathematik  at TUM as
John-von-Neumann 
professor.

\section{Definitions and main results}

\subsection{Fock space and second quantization}\label{SecondQ}
First we recall some basic facts. 
Let $\h$ be a Hilbert space. The Fock space over $\h$ is defined by
\[
 \Fock(\h)=\oplus_{n=0}^{\infty}\otimes_{\mathrm{s}}^n\h,
\]
where $\otimes_{\mathrm{s}}^n\h$ means the $n$-fold symmetric tensor
product of $\h$ with the convention $\otimes_{\mathrm{s}}^0\h=\BbbC$.
The vector $\Omega=1\oplus0\oplus\cdots\in\Fock(\h)$ is called the Fock
vacuum.

We denote by $a(f)$ the annihilation operator on $\Fock(\h)$ with test
vector $f\in\h$ \cite[Sec. X.7]{ReSi2}. By definition, $a(f)$ is densely
defined, closed, and antilinear in $f$. The adjoint $a(f)^*$ is the
adjoint of  $a(f)$ and  called the creation operator. Creation and
annihilation operators satisfy the canonical
commutation relations
\[
 [a(f),a(g)^*]=\la f,g\ra_{\h}\one,\ \ \ [a(f),a(g)]=0=[a(f)^*,a(g)^*]
\]
on the finite particle subspace
\[
 \Fock_0(\h)=\bigcup_{m=1}^\infty\{\varphi=\varphi_0\oplus\varphi_1\oplus\cdots
 \in\Fock(\h)\, |\, \varphi_n=0,\ \mathrm{for}\, n\ge m \},
\]
where $\la\cdot, \cdot\ra_{\h}$ denotes the inner product on $\h$ and
$\one$ denotes the identity operator. We  introduce a further  important
subspace of $\Fock(\h)$. Let  $\mathfrak{s}$ be a subspace of $\h$.
We define
\[
 \Ffin(\mathfrak{s})=\mathrm{Lin}\{a(f_1)^*\dots a(f_n)^*\Omega,\
 \Omega\, |\, f_1,\dots,f_n\in\mathfrak{s},\ n\in\BbbN\},
\]
where $\mathrm{Lin}\{\cdots\}$ means the linear span  of the  set
$\{\cdots\}$.
If $\mathfrak{s}$ is dense in $\h$, so is $\Ffin(\mathfrak{s})$ in $\Fock(\h)$.

Let $b$ be a contraction operator from $\h_1$ to $\h_2$, i.e., $\|b\|\le 1$.
The linear operator $\Gamma(b):\Fock(\h_1)\to \Fock(\h_2)$ is defined by
\[
 \Gamma(b)\upharpoonright\otimes_\mathrm{s}^n\h_1=\otimes^n b
\]
with the convention $\otimes^0 b=\one$.
It is well known that
\begin{align*}
\Gamma(b)a(f)^*=a(bf)^*\Gamma(b),\ \ 
\Gamma(b)a(b^*f)=a(f)\Gamma(b).
\end{align*}
For a densely defined closable operator $c$ on $\h$, $\dG(c):\Fock(\h)\to
\Fock(\h)$ is defined by
\[
 \dG(c)\upharpoonright \hotimes^n_{\mathrm{s}}\D(c)=\sum_{j=1}^{n}
\one\otimes\cdots\otimes\underset{j\,  \mathrm{th}}{c}\otimes \cdots\otimes\one
\]
and 
\[
 \dG(c)\Omega=0
\]
 where $\hotimes$ 
means the algebraic tensor product and for any linear operator $A$,
  $\D(A)$ denotes   the domain of $A$.
Here in the $j$-th summand $c$ is at the $j$-th entry.
Clearly $\dG(c)$ is closable and we denote its closure by the same
symbol. As an  example,  the number operator $N_{\mathrm{f}}$ is given
by  $N_{\mathrm{f}}=\dG(\one)$.

Let $\h_1$ and $\h_2$ be Hilbert spaces. Then there exists an isometry
$\mathsf{U}:\Fock(\h_1\oplus\h_2)\to \Fock(\h_1)\otimes\Fock(\h_2)$ such that 
\begin{align*}
\mathsf{U}\Omega&=\Omega\otimes\Omega,\\
\mathsf{U}a(h_1\oplus h_2)\mathsf{U}^*&=a(h_1)\otimes \one +\one \otimes a(h_2).
\end{align*}

\subsection{Definition of the Hamiltonian}

We consider  $N$ electrons with mass 
 $m_{\mathrm{e}}$ and charge $-e$, one nucleus with mass
 $m_{\mathrm{n}}$ and charge $Ze$,   moving  in   $3$-dimensional space
and coupled to   the quantized radiation field.
The  electrons are    fermions with spin $1/2$ and the nucleus is spinless.
 The Hilbert space
of state vectors is 
\[
 \Hil^{N+1}=L^2(\BbbR^3)\otimes [A_N\otimes^{N}L^2(\BbbR^3;\BbbC^2)] \otimes \mathcal{F},
\]
where  $\mathcal{F}$ is   the Fock space over $\oplus^{2}L^2(\BbbR^3)$,
\[
 \mathcal{F}=\Fock(\oplus^{2}L^2(\BbbR^3))
\]
and  $A_N$  denotes the  antisymmetrizer.
For $f\in L^2(\BbbR^3)$, we define $a_r(f)^{\#},\, r=1,2$, by 
\[
 a_r(f)^{\#}=a\Big(\oplus^2_{j=1}\delta_{rj}f\Big)^{\#},
\]
where $a^{\#}$ is either the creation or the annihilation operator on $\mathcal{F}$.
It is convenient to  use the notation $a_r(f)=\int_{\BbbR^3} \overline{f}(k)a_r(k)\,
\mathrm{d}k$.

The Hamiltonian of our system is the  Pauli-Fierz Hamiltonian  defined by
\begin{align}
H_{N}=&\frac{1}{2m_{\mathrm{n}}}\Big(-\im\nabla_0\otimes\one-Ze A(x_0)\Big)^2+
\sum_{j=1}^{N}\frac{1}{2m_{\mathrm{e}}}\Big\{\sigma_j\cdot\Big(-\im \nabla_j\otimes \one
+e A(x_j)\Big)\Big\}^2\nonumber\\
&+V\otimes \one +\one\otimes \Hf.\label{PFHami}
\end{align}
 Here the quantized vector potential
$A(x)=(A_{1}(x),A_{2}(x),A_3(x))$ is given  by

\begin{equation}
 A_{\mu}(x)=\sum_{r=1,2}\int_{\BbbR^3}\frac{\chi_{\sigma,\kappa}(k)}{\sqrt{2(2\pi)^3\omega(k)}}\Big\{ a_r(k)^*\ex^{-\im k\cdot
 x}\, + a_r(k)\, \ex^{\im k\cdot
 x}\Big\}e_{\mu}^{r}(k)\, \mathrm{d}k,\label{QVector}
\end{equation}
where   the form factor $\chi_{\sigma,\kappa}\
 (0\le\sigma<\kappa<\infty)$ is for simplicity choosen as 
$\chi_{\sigma,\kappa}=\chi_{\kappa}-\chi_{\sigma}$,
  $\chi_r$ with   the indicator function
 of the ball of  radius $r$.  $\sigma$ and $\kappa$ is  the 
 infrared cutoff and  ultraviolet cutoff, resp..
The polarization vector are denoted by $e^r=(e_1^r,e_2^r,e^r_3),\ r=1,2$.
Together with $k/|k|$  they form a basis, which for concreteness is taken
as
\[
 e^1(k)=\frac{(k_2,-k_1,0)}{\sqrt{k_1^2+k_2^2}},\ 
 e^2(k)=\frac{k}{|k|}\wedge e_1(k).
\]
Then
 $e^r(k)\cdot
e^s(k)=\delta_{rs}$ and $e^r(k)\cdot k=0$ a.e..
$\sigma_j=(\sigma_{j1}, \sigma_{j2}, \sigma_{j3})$ denotes the spin
matrix for the $j$-the particle.
The Hamiltonian of the free photon field  $\Hf$ is defined by
\begin{equation}
 \Hf=\dG(\oplus^{2}\omega),\ \ \omega(k)=|k|.\label{FreeH}
\end{equation}

We will prescribe  the following conditions for  $V$.
\begin{itemize}
\item[(V.1)] $V$ is a pair potential of the form
\begin{align*}
 V(x_0,\dots,x_N)=&\sum_{1\le i<j\le
 N}v(x_i-x_j)+\sum_{j=1}^Nw(x_0-x_j)\\
=:&\sum_{0\le i<j\le N}V_{ij}(x_i-x_j).
\end{align*}
Each $V_{ij}$  is  infinitesimally small with respect to $-\Delta$ in the sense that
 there exists sufficiently small $\varepsilon>0$ and
$b_{\varepsilon}>0$ such that 
\begin{eqnarray}
 \|V_{ij}f\|\le \vepsilon \|-\Delta f\|+b_{\vepsilon}\|f\|,\ \ f\in\D(-\Delta),\label{RelativeB}
\end{eqnarray}
where $-\Delta=-\sum_{j=0}^N\Delta_j$.
\item[(V.2)]  $v$ and $w$ are in $L^2_{\mathrm{loc}}(\BbbR^3)$.
      Moreover $V_{ij}(x)\to 0$ as $|x|\to \infty$.
\end{itemize}
As for the self-adjointness of $H_{N}$  the  following result is well-known.

\begin{Prop}{\rm\cite{Hiroshima2}}\label{SAPF} Assume (V.1).
 Then, for arbitrary $Z$, coupling $e$, mass $\emass, m_{\mathrm{n}}>0$ and cutoffs
 $\sigma,\kappa$ with $0\le \sigma<\kappa<\infty$,  $H_{N}$ is self-adjoint on $\D(-\Delta\otimes
 \one)\cap\D(\one\otimes \Hf)$ and bounded from below. Moreover $H_{N}$ is
 essentially self-adjoint on any core of $-\Delta\otimes \one +\one
 \otimes \Hf$.
\end{Prop}
\begin{rem}
{\rm
The proposition holds also for  massive photons, i.e., for the
 dispersion relation  $\omega_m(k)=\sqrt{k^2+m^2}$
 instead of $\omega(k)=|k|$.
The proof uses the  functional integral representation for
 $m>0$ as
established in \cite{Hiroshima3} and  is otherwise in essence identical to the one in
\cite{Hiroshima2}.
}
\end{rem}

Let $\Pt$ be the total momentum operator, namely
\[
 \Pt=-\im\sum_{j=0}^N \nabla_j\otimes \one+\one\otimes \Pf,
\]
where $\Pf=(P_{\mathrm{f},1},
P_{\mathrm{f},2},P_{\mathrm{f},3})=(\dG(\oplus^2k_1),\dG(\oplus^2k_2),\dG(\oplus^2k_3))$
is 
the   momentum operator of the electromagnetic field.
Each component $P_{\mathrm{tot},j},\ j=1,2,3$ of $\Pt$ is essentially self-adjoint. We denote its
closure by the same symbol $P_{\mathrm{tot},j}$.
To obtain $H(P)$ in (\ref{QHamiltonian}), the Hamiltonian at fixed total
momentum $P$, formally we regard $\Pt=P$ as a parameter and simply  substitute in
(\ref{PFHami}) as
\[
-\im\nabla_0\otimes\one=P+\im\sum_{j=1}^N\nabla_j\otimes\one-\one\otimes\Pf.
\]
In the resulting Hamiltonian we may then set $x_0=0$.
To be more precise, let us define, for all $x_0\in\BbbR^3$, 
\[
 W(x_0)=\exp\{\im x_0\cdot(\Pt+\im\nabla_0\otimes \one)\}
\]
acting on $\Hil^N$.  Since
$x_0\to W(x_0)$ is strongly continuous, we can define the  fiber direct
integral operator
\[
 W=\int^{\oplus}_{\BbbR^3}W(x_0)\, \mathrm{d}x_0
\]
acting on $\Hil^{N+1}=\int^{\oplus}_{\BbbR^3}\Hil^N\, \mathrm{d}x_0$,
where
\[
 \Hil^N=[A_N\otimes^NL^2(\BbbR^3;\BbbC^2)]\otimes\mathcal{F}.
\]
Let $\mathcal{U}$ be the Fourier transformation with respect to the
variable $x_0$, acting in $L^2(\BbbR^3)\otimes [A_N\otimes^N L^2(\BbbR^{3};\BbbC^2)]$,
\[
 (\mathcal{U}f)(P,x_1,\dots,x_{N})=(2\pi)^{-3/2}\int_{\BbbR^3}\ex^{-\im
 P \cdot x_0}f(x_0,x_1,\dots,x_N)\, \mathrm{d}x_0.
\]
The linear operator $U_F=\mathcal{U}\otimes \one$ is unitary on $\Hil^{N+1}$.
Next we define a unitary operator on $\Hil^{N+1}$ by $U=U_FW$. The
unitary operator $U$ induces the identification of 
$\Hil^{N+1}$ with  $\int^{\oplus}_{\BbbR^3}\Hil^N\, \mathrm{d}P$, which is
concretely given by
\begin{align*}
 &(U\psi)^{(n)}(P,x_1,\dots,x_N,k_1,\dots,k_n)\\
=&(2\pi)^{-3/2}\int_{\BbbR^3}\ex^{-\im
 (P-\sum_{j=1}^nk_j)\cdot x_0}\psi^{(n)}(x_0,x_1-x_0,\dots, x_N-x_0,k_1,\dots,k_n)\, \mathrm{d}x_0
\end{align*}
for $\psi=\oplus_{n=0}^{\infty}\psi^{(n)}\in\Hil^{N+1}$.
 It is not hard
to check that
\[
 U P_{\mathrm{tot},j}U^*=\int^{\oplus}_{\BbbR^3}P_j\, \mathrm{d}P.
\] 
Hence the operator $U$ provides the   direct integral decomposition of $\Hil^{N+1}$
with respect to the value of the total momentum.  It  can be  easily seen that 
$\ex^{\im\lambda\cdot \Pt}H_{N}\subseteq H_{N}\, 
\ex^{\im\lambda\cdot \Pt}$ 
for all $\lambda\in\BbbR^3$, i.e.,
$\Pt$ and $H_{N}$ strongly commute. Thus $UH_{N}U^*$ is a decomposable
operator, i.e., $UH_{N}U^*$ can be represented by  the fiber direct
integral 
\begin{align}
 UH_{N}U^*=\int^{\oplus}_{\BbbR^3}\mathscr{H}(P)\, \mathrm{d}P.\label{DirInt}
\end{align}
Clearly  $\mathscr{H}(P)$ is a self-adjoint operator  for a.e. $P$ acting
in $\Hil^N$.

We introduce a dense subspace of $\Hil^{N}$ by
\[
 \Hil_{\mathrm{fin}}^{N}=[A_N\hat{\otimes}^{N} C_0^{\infty}(\BbbR^{3}_x;\BbbC^2)]
\hat{\otimes}\,  \Ffin(\oplus^{2}C_0^{\infty}(\BbbR^3)).
\]
On $\Hil_{\mathrm{fin}}^{N}$  we can write down $\mathscr{H}(P)$ as follows,
\begin{align}
\mathscr{H}(P)=&\sum_{j=1}^{N}\frac{1}{2m_{\mathrm{e}}}\Big\{\sigma_j\cdot\Big(-\im \nabla_j\otimes \one
+e
 A(x_j)\Big)\Big\}^2\nonumber\\
&+ \frac{1}{2m_{\mathrm{n}}}\Big(P+\im\sum_{j=1}^{N}\nabla_j\otimes
 \one-\one \otimes\Pf-Ze A(0)\Big)^2\nonumber\\
&+\tilde{V}\otimes \one+\one\otimes \Hf, \label{Hami}
\end{align}
where 
\begin{align}
 \tilde{V}(x_1,\dots,x_{N})=\sum_{1\le
 i<j\le N}v(x_i-x_j)+\sum_{j=1}^{N}w(x_j).\label{Potential}
\end{align}
The   symmetric operator  $H(P)$ is now defined  by the right hand side of
(\ref{Hami}).
Clearly $H(P)$ is closable and we denote its closure by the same symbol.
Note that,  by (\ref{Hami}),
\[
	\mathscr{H}(P)=H(P)
\]
 on the dense subspace
$\Hil_{\mathrm{fin}}^N$.

\subsection{Main results}

Our first result  concerns    the self-adjointness of $H(P)$.

\begin{Thm}\label{SA} Assume (V.1).
For  arbitrary  $Z$, coupling $e$, cutoffs $\sigma,\kappa$ with
 $0\le \sigma<\kappa<\infty$ and  total momentum $P$, $H(P)$ is
 self-adjoint on
 $\cap_{j=1}^{N}\D(-\Delta_j\otimes\one)\cap\D(\one\otimes\Pf^2)\cap\D(\one\otimes\Hf)$
 and essentially
 self-adjoint on any core of
 $-\sum_{j=1}^{N}\Delta_j\otimes\one+\one\otimes \Pf^2+\one\otimes
 \Hf$.
In particular, $H(P)$ is essentially self-adjoint on
 $\Hil_{\mathrm{fin}}^{N}$. Moreover
\[
 UH_NU^*=\int^{\oplus}_{\BbbR^3}H(P)\, \mathrm{d}P.
\]
\end{Thm}

We introduce the energy inequality and the binding condition.

Let $H_m(P)$ be the Hamiltonian (\ref{Hami})  with the photon
dispersion
relation $\omega_m(k)=\sqrt{k^2+m^2}$ and $E_m(P)$ be the infinimum of the
spectrum of $H_m(P)$, i.e.,

\[
E_m(P)=\inf\mathrm{spec}(H_m(P))
\]
with $\mathrm{spec}(A)$ denoting the spectrum of the linear operator
$A$. The {\it energy inequality} reads
\[
 E_m(0)\le E_m(P)\tag{E.I.}
\]
for any sufficiently small $m\ge 0$. As shorthand we set $H_0(P) = H(P)$
and let $E_0(P)= E(P)$.

Let $\mathcal{D}_R=\{\vphi\in\Hil_{\mathrm{fin}}^N\,|\, \vphi(x)=0\
\mbox{for   $|x|<R$}\}$ and introduce a threshold energy $\Sigma(P)$
by
\begin{align}
 \Sigma(P)=\lim_{R\to\infty}\Big(\inf_{\vphi\in\mathcal{D}_R,\|\vphi\|=1}
\la\vphi, H(P)\vphi\ra\Big).\label{Threshold}
\end{align}
The {\it binding condition} for our model is  stated as
\begin{align*}
  \Sigma(P)> E(P).\tag{B.C.}
\end{align*}
In case of vanishing coupling to the Maxwell field the binding condition
reduces to more standard versions  based on cluster decomposition, as will
be explained in Appendix D.
We note that the binding condition  depends on the parameter $P$. 
Let $\Lambda $ be the  set on which   the binding condition is satisfied, i.e.,
\[
 \Lambda=\{P\in\BbbR^3\, |\, \Sigma(P)> E(P)\}.
\]

First we treat neutral atoms. Mathematically the
neutrality condition is expressed as 
\begin{align*}
 N=Z.\tag{N}
\end{align*}

\begin{Thm}\label{ExMassless}
Assume  (V.1), (V.2), (E.I.),  (N), and  the infrared cutoff
 $\sigma=0$.
 If $P\in\Lambda$ and $|P|< \Nm$, then $H(P)$ has a ground state.
\end{Thm}

The condition $P\in\Lambda$ is implicit.
But it  can be written  more explicitly  under  stronger assumptions.

Let $\Pi_N$ be the set of the subsets of $\{0,1,2,\dots,N\}$. We denote by
$H_{\beta}$ the Hamiltonian of the form (\ref{PFHami}), but only
refering to  the
particles in the set $\beta\in\Pi_N$, i.e., 
\begin{align*}
 H_{\beta}=&\frac{1}{2m_{\mathrm{n}}}\Big(-\im\nabla_0\otimes\one-Ze A(x_0)\Big)^2
+\sum_{j\in\beta\backslash
 \{0\}}\frac{1}{2m_{\mathrm{e}}}\Big\{\sigma_j\cdot \Big(-\im\nabla_j\otimes\one
 +e A(x_j)\Big)\Big\}^2\\&+\sum_{i,j\in\beta,  0\le i<j\le N}V_{ij}\otimes\one
 +\one\otimes \Hf, 
\end{align*}
if $0\in\beta$, and, if $0\notin \beta$,
\begin{align*}
 H_{\beta}=&
\sum_{j\in\beta}\frac{1}{2m_{\mathrm{e}}}\Big\{\sigma_j\cdot\Big(-\im\nabla_j\otimes\one
 +e A(x_j)\Big)\Big\}^2\\&+\sum_{i,j\in\beta,0\le i<j\le N}V_{ij}\otimes\one
 +\one\otimes \Hf.
\end{align*}
Let us  introduce
\[
 E_{\beta}=\inf \mathrm{spec}(H_{\beta}) \ ,
\]
and let $E_{N}=\inf\mathrm{spec}(H_{N})$. (With  our  notation, $E_{N}=E_{\{0,1,\dots,N\}}$.)
The binding energy for  the  Hamiltonian $H_{N}$ is defined by
\[
 E_{\mathrm{bin}}=\min\big\{E_{\beta}+E_{\bar{\beta}}\, |\,
 \beta\in\Pi_N\, \mathrm{and}\, \beta\neq\emptyset, \{0,1,\dots,N\}\big\}-E_{N},
\]
where $\bar{\beta}$ denotes the complement of $\beta$.

\begin{Thm}\label{MainTheorem}Assume  (V.1), (V.2), (E.I.), (N), and  
the infrared cutoff $\sigma=0$.  If   $E_{\mathrm{bin}}>0$  and 
\[
 |P|<\min\big\{\Nm,\sqrt{2\Nm
 E_{\mathrm{bin}}}\big\},
\]
then $H(P)$ has a ground state.
\end{Thm}

As explained  before, for ions  we need an infrared cutoff.

\begin{Thm}\label{Iontype}
Assume  (V.1), (V.2), (E.I.),  and  a non-neutral system, i.e., 
 (N) does not hold. Suppose that $\sigma>0$. 
\begin{itemize}
\item[{\rm (i)}] If $P\in\Lambda$ and $|P|<\Nm$,  then $H(P)$ has a ground
		 state.
\item[{\rm (ii)}]  If
 $E_{\mathrm{bin}}>0$ and $|P|<\min\{\Nm,\sqrt{2\Nm E_{\mathrm{bin}}}\}$,
 then $H(P)$ has a ground state. 
\end{itemize}
\end{Thm}

To establish the   parameter values for which  the binding condition  holds
  is a  difficult problem.
Indeed,  to prove $E_{\mathrm{bin}}>0$  in case of a  fixed nucleus is already
 very hard work \cite{LL}. 
Thus the reader might worry whether the binding condition can be satisfied at
  all. We will prove it for  $\Nm$ sufficiently large.

In the limit $\Nm\to \infty$, $H_N$ of (\ref{PFHami}) converges to
$H_N^{\infty}$ defined by
\[
 H_N^{\infty}=\sum_{j=1}^{N}\frac{1}{2\emass}\Big\{\sigma_j\cdot\Big(-\im \nabla_j\otimes \one
+e A(x_j)\Big)\Big\}^2+\tilde{V}\otimes \one +\one\otimes \Hf, \label{Static}
\]
where we have set $x_0=0$ and $\tilde{V}$ is defined in (\ref{Potential}). Let $E_N^{\infty}=\inf
\mathrm{spec}(H_N^{\infty})$.
With the cluster  decomposition from above the binding energy for
$H_N^{\infty}$ is given by
\[
 E_{\mathrm{bin}}^{\infty}=\min\big\{E_{\beta}^{\infty}+E_{\bar{\beta}}^{\infty}\, |\,
 \beta\subset \{1,\dots, N\}\, \mathrm{and}\, \beta\neq\emptyset,
 \{1,\dots,N\}\big\}-E_{ N}^{\infty}.
\]
In \cite{LL}  conditions are provided under which $E_{\mathrm{bin}}^{\infty}>0$.

\begin{rem}
{\rm
In  \cite{LL} $E_{\mathrm{ bin}}^{\infty}>0$ is proved for
 molecules and atoms  with a
 smooth cutoff function $\hat{\vphi}$ instead of the sharp cutoff
 $\chi_{0,\kappa}$ used here.
There is no difficulty in extending our main results to a smooth cutoff.
}
\end{rem}

\begin{Prop}\label{PFBinding} Assume (V.1), (V.2) and (E.I.).
For sufficiently large $\Nm$,
the binding condition (B.C.) holds provided $|P|<\sqrt{\Nm
 E_{\mathrm{bin}}^{\infty}}$.
\end{Prop}
{\it Proof.} See Appendix \ref{PrBC}. $\Box$\medskip\\

We remark  that Proposition \ref{PFBinding} is needed as an input for
 Theorem \ref{Iontype}.

\section{Proof of Theorem \ref{SA}}

Theorem \ref{SA} is proved in using the following strategy.  Firstly we
define  a new Hamiltonian $\tilde{H}(P)$
which is self-adjoint and which coincides with $H(P)$ on a dense domain.
Secondly we prove
\begin{equation}\label{aa}
UH_NU^*=\int^{\oplus}_{\BbbR^3}\tilde{H}(P)\, \mathrm{d}P
\end{equation}
and
clarify  the domain and the domain of essential self-adjointness of $\tilde{H}(P)$ by applying Proposition \ref{SAPF} and (\ref{aa}).
Finally we show that this self-adjoint operator equals  $H(P)$.
Clearly, the essential point lies in the choice of
$\tilde{H}(P)$.
The reader might think that  the simplest way to define a new Hamiltonian
$\tilde{H}(P)$ is by just taking the Friedrichs
extension $H_1(P)$ of $H(P)$. However,  in this case, it seems difficult to  establish
the measurability of $H_1(P)$ in the sense that the map $P\to \la
\varphi,(H_1(P)+\im)^{-1} \psi \ra$ is measurable. On the other
hand,
 the measurability of $\tilde{H}(P)$ is required to define
 $\int_{\BbbR^3}^{\oplus}\tilde{H}(P)\, \mathrm{d}P$. Therefore we will
adopt another  construction for the  Hamiltonian $\tilde{H}(P)$.
We will see that the construction  of  the Hamiltonian $\tilde{H}(P)$
 which will put to use in Section \ref{NoSS}.

\subsection{Definitions}\label{Definitions}
 
Let 
\[
 H_A=\frac{1}{2\Nm}\Big(-\im\nabla_1\otimes \one-Ze
 A(-x_1)\Big)^2+\frac{1}{2}\one\otimes \Hf.
\]
By Proposition \ref{SAPF}, $H_A$ is self-adjoint on
$\D(-\Delta_1\otimes\one)\cap\D(\one\otimes \Hf)$ for all $e$ and
cutoffs.
For all $P\in\BbbR^3$,  let
$\mathscr{V}(P)$ be a unitary operator defined by
\begin{align}
\mathscr{V}(P)=\exp\Big\{\im x_1\cdot\Big(P+\im\sum_{j=2}^{N}\nabla_j
\otimes\one-\one\otimes\Pf\Big)\Big\}.\label{VofP}
\end{align}
  We introduce $K(P)$ by
\begin{align}
 K(P)=\mathscr{V}(P)H_A\mathscr{V}(P)^*,\label{KofP}
\end{align}
then $K(P)$ is also self-adjoint  for all $e$ and $P\in\BbbR^3$, and
\[
 K(P)\Psi=\frac{1}{2\Nm}\Big(P+\im\sum_{j=1}^{N}\nabla_j\otimes
 \one-\one \otimes\Pf-Ze A(0)\Big)^2\Psi+\frac{1}{2}\one\otimes \Hf\Psi
\]
for  $\Psi\in\Hil_{\mathrm{fin}}^{N}$.

Let 
\[
 H_{\mathrm{PF}}=\sum_{j=1}^{N}\frac{1}{2\emass}\Big\{\sigma_j\cdot\Big(-\im \nabla_j\otimes \one
+e A(x_j)\Big)\Big\}^2+\tilde{V}\otimes \one +\frac{1}{2}\one\otimes \Hf
\]
acting in $\Hil^{N}$. 
 By (V.1), $\tilde{V}$ is infinitesimally small with respect to
 $-\sum_{j=1}^N\Delta_j$. 
Hence, by  Proposition \ref{SAPF}, $H_{\mathrm{PF}}$
 is self-adjoint on
 $\cap_{j=1}^{N}\D(-\Delta_j\otimes\one)\cap\D(\one\otimes\Hf)$,
essentially self-adjoint on $\Hil^{N}_{\mathrm{fin}}$ for arbitrary
 coupling and cutoffs.

Now we define a densely defined symmetric form $\mathsf{s}_{P}$ as
follows
\begin{align*}
Q(\mathsf{s}_P)&=\D(|H_{\mathrm{PF}}|^{1/2})\cap\D(K(P)^{1/2}),\ \
 \mbox{(form domain)}\\
 \mathsf{s}_P(\vphi,\psi)&=\la
 |\hat{H}_{\mathrm{PF}}|^{1/2}\vphi,|\hat{H}_{\mathrm{PF}}|^{1/2}\psi\ra+\la
 K(P)^{1/2}\vphi,K(P)^{1/2}\psi\ra\\
&\hspace{1.5cm}+\inf\mathrm{spec}(H_{\mathrm{PF}})\la\vphi,\psi\ra,
\end{align*}
 for $\vphi,\psi\in Q(\mathsf{s}_P)$, where
$\hat{A}=A-\inf\mathrm{spec}(A)$. $\mathsf{s}_P$ is closed and 
semibounded. Let $\tilde{H}(P)$ be the self-adjoint operator associated with
$\mathsf{s}_P$. Then $\tilde{H}(P)$ is a self-adjoint extension of
$H_{\mathrm{PF}}+K(P)$ and the formula 
\begin{align*}
\tilde{H}(P)\Psi=H(P)\Psi
\end{align*}
holds for all $\Psi\in\Hil_{\mathrm{fin}}^{N}$. 

\begin{lemm}\label{measurability}
The mapping $P\to (\tilde{H}(P)+\im)^{-1}$ is measurable, i.e., for all
 $\vphi,\psi\in\Hil^{N}$, $P\to \la\vphi,(\tilde{H}(P)+\im)^{-1}\psi\ra$ is
 a measurable mapping.
\end{lemm}
$Proof.$\ \ By Kato's strong Trotter product formula \cite[Theorem
 S.21]{ReSi1},
 we have
\begin{eqnarray}
\ex^{-t\tilde{H}(P)}=\slim\Big(\ex^{-tH_{\mathrm{PF}}/n}\ex^{-tK(P)/n}\Big)^n.\label{StrKato}
\end{eqnarray}
Since $P\to \ex^{-sK(P)}=\mathscr{V}(P)\ex^{-sH_A}\mathscr{V}(P)^*$ is strongly continuous,
$P\to\ex^{-t\tilde{H}(P)}$ is measurable by (\ref{StrKato}). Therefore, we
obtain the desired assertion. $\Box$ 

\hspace{5mm}

Thanks to the above lemma and \cite[Theorem XIII.85]{ReSi4}, one can
define a self-adjoint operator $H'$ on $\Hil^N$ by
\[
 H'=\int^{\oplus}_{\BbbR^3}\tilde{H}(P)\, \mathrm{d}P.
\]

\begin{Prop}\label{FibreH}
\[
UH_{N}U^*=\int^{\oplus}_{\BbbR^3}\tilde{H}(P)\, \mathrm{d}P.
\]
\end{Prop}
To prove this we need some preparations.
Let
\begin{align}
 L=-\sum_{j=1}^{N}\Delta_j\otimes\one+\Big(k\otimes\one+\im\sum_{j=1}^{N}\nabla_j\otimes\one-\one\otimes\Pf\Big)^2+\one\otimes\Hf.\label{DefL}
\end{align}
$L$ is closable and we denote its closure by the same symbol.

\begin{lemm}
$L$ is essentially self-adjoint on 
\begin{align}
 \mathcal{V}=\big[A_N\hat{\otimes}^{N}C_0^{\infty}(\BbbR^3_x;\BbbC^2)\big]
\hat{\otimes}\, C_0^{\infty}(\BbbR^3_k)\, \hat{\otimes}\, \Ffin(\oplus^{2}C_0^{\infty}(\BbbR^3)).\label{subspaceV}
\end{align}
and 
\begin{align}
 L=U(-\Delta\otimes\one+\one\otimes\Hf)U^*. \label{TestOp}
\end{align}
\end{lemm}
{\it Proof.}\ \   Essential self-adjointness of $L$ on $\mathcal{V}$ is  proven by  Nelson's
commutator theorem \cite[Theorem X.37]{ReSi2} with a test operator
$J=-\sum_{j=1}^N\Delta_j\otimes\one+k^2\otimes\one+\one\otimes
\Pf^2+\one\otimes\Hf+\one\otimes\one$. 
We can confirm that (\ref{TestOp}) holds on $\mathcal{V}$. Since
$\mathcal{V}$ is a core of $L$, we conclude (\ref{TestOp}) as an
operator equality.
$\Box$
\medskip\\
{\sf Proof of Proposition \ref{FibreH}}\\
By Proposition \ref{SAPF} and the above lemma, $UH_{N}U^*$ is essentially
self-adjoint on $\mathcal{V}$.
On $\mathcal{V}$ we can  check that $UH_{N}U^*=H'$ which implies
the proposition. $\Box$

\subsection{Domain of self-adjointness for  $H(P)$}

We prove Theorem \ref{SA} by series of lemmata. The first lemma is a simple
application of the closed graph theorem.
\begin{lemm}\label{EssSA1}
Let $A$ and $B$ be self-adjoint operators. Suppose that  $\D(A)=\D(B)$.
 Then there exists $C_1>0$ and $C_2>0$ such that
\[
 C_1\|\vphi\|_A\le \|\vphi\|_B\le C_2\|\vphi\|_A,
\]
where, for a linear operator $T$,
 $\|\vphi\|_T^2=\|T\vphi\|^2+\|\vphi\|^2$ for $\vphi\in \D(T)$.
In particular, $A$ is essentially self-adjoint on any core of $B$ and
 $B$ is essentially self-adjoint on any core of $A$.
\end{lemm}
$Proof$.\ \ Let $\mathcal{D}=\D(A)=\D(B)$. Norm spaces
$D_A=(\mathcal{D},\|\cdot\|_A)$ and $D_B=(\mathcal{D},\|\cdot\|_B)$
are both closed  by the
self-adjointness of $A$ and $B$. Now let $i: D_A\to D_B$ defined by
\[
 i\vphi=\vphi,\ \ \vphi\in D_A.
\]
Then the graph of $i$ is closed. Indeed let
\[
 \mathrm{gr}(i)=\{\vphi\oplus i\vphi\, |\, \vphi\in
 \mathcal{D}\}\subseteq D_A\oplus D_B
\]
and let $\{\vphi_n\oplus i\vphi_n\}$ be a Cauchy sequence in
$\mathrm{gr}(i)$. Then $\{\vphi_n\}$ is also Cauchy in $D_A$, $D_B$ and
the underlying Hilbert space. Thus there exists
$\vphi=\lim_{n\to\infty}\vphi_n\in \mathcal{D}$,
$\lim_{n\to\infty}A\vphi_n=A\vphi$ and
$\lim_{n\to\infty}B\vphi_n=B\vphi$ by the closedness of  self-adjoint
operators. Therefore $\{\vphi_n\oplus i\vphi_n\}$ is a convergent
sequence in $\mathrm{gr}(i)$.
Applying the closed graph theorem, $i$ is bounded and 
\[
 \|\vphi\|_B\le C \|\vphi\|_A
\]
 for some constant $C>0$. From this $B$ is essentially self-adjoint on
 any core of $A$. Interchanging the role of $A$ and $B$, we also
 conclude the remaining assertion. $\Box$

\begin{lemm}\label{EssSA2}
Let $A$ and $B$ be positive decomposable operators on the Hilbert space 
$\int_M^{\oplus}\mathcal{X}\, \mathrm{d}\mu(m)$ with $\D(A)=\D(B)$.
 Then  $\D(A(m))=\D(B(m))$, furthermore the self-adjoint operator
$A(m)$ is essentially self-adjoint  on any core of $B(m)$, and  the
 self-adjoint operator $B(m)$ is
  essentially self-adjoint on any core of $\D(A(m))$ for $\mu$-a.e. $m$.
\end{lemm}
$Proof$.\ \ By Lemma \ref{EssSA1}, there is a constant $d>0 $ so that 
\[
 \|A\vphi\|\le d(\|B\vphi\|+\|\vphi\|),\ \ \vphi\in\D(A).
\] 
Hence $C:=A(B+\one)^{-1}$ is a  bounded operator.

Since $A$ and $(B+\one)^{-1}$ are both  decomposable,
 $C$ is  also decomposable. Therefore we can
represent $C$ as $C=\int^{\oplus}_M C(m)\, \mathrm{d}\mu(m)$. 
Moreover 
it is not
hard to check that $C(m)=A(m)(B(m)+\one)^{-1}$ for $\mu$-a.e. $m$.
(Note that $A(m)$ and $B(m)$ are both self-adjoint for $\mu$-a.e..)
Hence, $A(m)(B(m)+\one)^{-1}$ is bounded and 
\[
 \|A(m)(B(m)+\one)^{-1}\| \le \|C\|
\]
for $\mu$-a.e.. Thus $A(m)\, \ex^{-tB(m)}=A(m)(B(m)+\one)^{-1}(B(m)+\one)\,
\ex^{-tB(m)}$ is bounded for all $t>0$.
This means $\ex^{-tB(m)}\D(A(m))\subseteq\D(A(m))$ for all $t>0$. By applying \cite[Theorem
X.49]{ReSi2}, $B(m)$ is essentially self-adjoint on $\D(A(m))$.
Similarly $A(m)$ is essentially self-adjoint on $\D(B(m))$ for
$\mu$-a.e..
Therefore $\D(A(m))=\D(B(m))$ and we have the desired result by Lemma \ref{EssSA1}.
$\Box$

\hspace{5mm}

\begin{lemm}\label{Domain}
Let 
\[
 L(P)=-\sum_{j=1}^{N}\Delta_j\otimes\one+\overline{\Big(P+\im\sum_{j=1}^{N}\nabla_j\otimes\one-\one\otimes\Pf\Big)^2}+\one\otimes\Hf.
\]
acting in $\Hil^N$. Then, for all $P\in\BbbR^3$, $L(P)$ is self-adjoint
 on $\cap_{j=1}^N\D(-\Delta_j\otimes\one)\cap \D(\one\otimes\Pf^2)\cap
 \D(\one\otimes\Hf)$,
essentially self-ajoint on $\Hil_{\mathrm{fin}}^N$. Moreover
\begin{align}
 L=\int^{\oplus}_{\BbbR^3}L(P)\, \mathrm{d}P.\label{Fibre}
\end{align}
\end{lemm}
{\it Proof.}\ \  By the functional
calculus, we   confirm that 
$\D(L(P))=\cap_{j=1}^N\D(\Delta_j\otimes\one)\cap
\D(\one\otimes\Pf^2)\cap\D(\one\otimes\Hf)$. Thus 
by applying Lemma \ref{EssSA1}, $L(P)$ is essentially self-adjoint on
$\Hil_{\mathrm{fin}}^N$. On the subspace $\mathcal{V}$ defined  by
(\ref{subspaceV}) one can
easily see (\ref{Fibre}). Thus we  conclude
(\ref{Fibre}) as an operator equality. $\Box$

\begin{lemm}\label{IndPIneq}
Let $\tilde{H}^{V=0}(P)$ be the Hamiltonian $\tilde{H}(P)$ with $V=0$.
Then, for all $P\in\BbbR^3$, there is a finite constant $C>0$
independent of $P$ such that
\[
 \|\tilde{H}^{V=0}(P)\vphi\|\le C\big(\|\tilde{H}(P)\vphi\|+\|\vphi\|\big),\ \ \vphi\in\Hil_{\mathrm{fin}}^N.
\]
\end{lemm}
{\it Proof.}\ \ Let $H^{V=0}_N$ be the Hamiltonian $H_N$ with $V=0$.
By Proposition \ref{SAPF}, two self-adjoint operators $H^{V=0}_N$ and
$H_N$ have the same domain. Hence there is a constant $C>0$ such that 
\[
  \|UH^{V=0}_NU^*\Psi\|\le C\big(\|UH_NU^*\Psi\|+\|\Psi\|\big)
\]
by Lemma \ref{EssSA1}. 
Let  $M_n(P)=\{k\in\BbbR^3\, |\, |k_j-P_j|\le\frac{1}{2n},\, j=1,2,3\}$
for $P\in\BbbR^3$.
Taking
$\Psi(k,x_1,\dots,x_N)=\eta_n(k)\vphi(x_1,\dots,x_N)$ with
$\eta_n=n^{3/2}\chi_{M_n(P)}$ 
 and $\vphi\in\Hil_{\mathrm{fin}}^N$, one has 
\begin{align*}
&\Big(\int_{\BbbR^3}|\eta_n(k)|^2\|\tilde{H}^{V=0}(k)\vphi\|^2\, \mathrm{d}k\Big)^{1/2}\\
\le& C\Big(\int_{\BbbR^3}|\eta_n(k)|^2\|\tilde{H}(k)\vphi\|^2\,
 \mathrm{d}k\Big)^{1/2}+C \|\vphi\|
\end{align*}
by Proposition \ref{FibreH}.
 Noting  that $k\to \tilde{H}^{V=0}(k)\vphi$ and
$k\to \tilde{H}(k)\vphi$ are strongly continuous, we can conclude the
 assertion by taking $n\to\infty$.  $\Box$ 
\medskip\\
\textsf{Proof of Theorem \ref{SA}}\\
By Proposition \ref{SAPF} and (\ref{TestOp}),  $UH_NU^*$ is
self-adjoint on $\D(L)$. By applying Lemma \ref{EssSA2} and  \ref{Domain},
$\tilde{H}(P)$ is self-adjoint on $\D(L(P))=\cap_{j=1}^N\D(\Delta_j\otimes\one)
\cap\D(\one\otimes\Pf^2)\cap\D(\one\otimes\Hf)$ and essentially
self-adjoint on $\Hil_{\mathrm{fin}}^N$ for $P\in\BbbR^3\backslash
\mathcal{N}$ where $\mathcal{N}$ is a measure zero set.

Let  $P_0\in\mathcal{N}$. We introduce a linear operator $\delta_P
\tilde{H}(P_0)$ by
\[
\delta_P \tilde{H}(P_0)= \tilde{H}(P)-\tilde{H}(P_0).
\]
For each $\Psi\in \Hil_{\mathrm{fin}}^{N}$ and $P\notin
\mathcal{N}$,
\begin{align*}
\delta_P \tilde{H}(P_0)\Psi&=[\tilde{H}(P_0)-\tilde{H}(P)]\Psi\\
&=\frac{1}{2\Nm}[2(P-P_0)\cdot (X-P)+3P^2-P^2_0-2P_0\cdot P]\Psi,
\end{align*}
where  $X=\im\sum_{j=1}^{N}\nabla_j\otimes \one-\one \otimes \Pf-Ze A(0)$.
We prove that there is a constant $C$ independent of $P$ and
$B_{P,P_0}>0$ which is finite for all $P\notin \mathcal{N}$ such that 
\begin{align}
 \|\delta_p\tilde{H}(P_0)\Phi\|\le C|P-P_0|\big(\|\tilde{H}(P)\Phi\|+B_{P,P_0}\|\Phi\|\big)\label{guhe}
\end{align}
for all $\Phi\in\D(\tilde{H}(P))$. For $\Psi\in
\Hil_{\mathrm{fin}}^{N}$ and $j=1,2,3$,
\begin{align*}
\|(X_j-P_j)\Psi\|
&  \le C_1\big(\|\tilde{H}^{V=0}(P)\Psi\|+
 \|\Psi\|\big)\\
& \le C_2\big(\|\tilde{H}(P)\Psi\|+
 \|\Psi\|\big)
\end{align*}
by Lemma \ref{IndPIneq}. Note that $C_2$ does not depend
on $P$. 
From this, we obtain (\ref{guhe}) for $\Phi\in\Hil_{\mathrm{fin}}^{N}$. 
Since $\Hil_{\mathrm{fin}}^{N}$ is a core of $\tilde{H}(P)$, we can
 extend the result to $\D(\tilde{H}(P))$.

Since $\mathcal{N}$ has measure zero, there is a $P\in\BbbR^3\backslash
\mathcal{N}$ such that $|P-P_0|C<1$. Thus, by (\ref{guhe}) and
the Kato-Rellich theorem \cite[Theorem X.12]{ReSi2},
$\tilde{H}(P_0)=\tilde{H}(P)+\delta_P\tilde{H}(P_0)$ is self-adjoint on $\D(\tilde{H}(P))=\D(L(P))$
and essentially self-adjoint on any core of $\tilde{H}(P)$.
Since, for all $P\in\BbbR^3$,  $\tilde{H}(P)$ is essentially
self-adjoint on  $\Hil_{\mathrm{fin}}^N$
and $H(P)\Psi=\tilde{H}(P)\Psi$ for $ \Psi\in\Hil_{\mathrm{fin}}^N$, we
have  $H(P)=\tilde{H}(P)$ for all $P$. 
$\Box$

\section{Properties of the ground state energy}

Let $H_{N,m}$ be the  Hamiltonian  (\ref{PFHami}) with the photon dispersion relation
$\omega_m(k)=\sqrt{k^2+m^2}$ instead of $\omega(k)=|k|$.
Note that Theorem \ref{SAPF} also  holds for $H_{N,m}$ with arbitrary
$m\ge 0$.
Therefore  $H_m(P)$ is self-adjoint on
$\cap_{j=1}^{N}\D(-\Delta_j\otimes\one)\cap\D(\one\otimes\Pf^2)\cap\D(\one\otimes
H_{\mathrm{f},m})$, essentially
 self-adjoint on $\Hil_{\mathrm{fin}}^N$ for all $e,m, Z$, cutoffs  and $P$
under the assumptions (V. 1) and (V. 2).
We denote the infinimum of the spectrum of $H_{N,m}$ 
by $E_{N,m}$.
The purpose of this section is to prove a few simple properties of the function $E_m(P)$.

\begin{Thm}\label{IneqGSE1}
Assume  (V.1), (V.2) and (E.I.).
For all $\Mass\ge 0$, $Z$,  coupling $e$, and cutoffs $0\le \sigma
 <\kappa<\infty$, the following assertions  hold.
\begin{itemize}
\item[{\rm (i)}]  The function $f(P)=\frac{1}{2\Nm}P^2-E_m(P)$ is a
		 convex function. In particular, $f(P)$ and hence $E_m(P)$ are  continuous in $P$. 
\item[{\rm (ii)}] For all $P\in\BbbR^3$,
\[
 E_m(P)-E_m(0)\le\frac{1}{2\Nm}P^2.
\]
\item[{\rm (iii)}] 
\[
 E_m(P-k)-E_m(P)\ge \begin{cases}
-\frac{|k||P|}{\Nm}+\frac{k^2}{2\Nm} &\mbox{if $|k|\le |P|$,}\\
-\frac{P^2}{2\Nm} & \mbox{if $|k|\ge |P|$ }
\end{cases}.
\]
\item[{\rm (iv)}] $E_m(0)=E_{N,\Mass}$.
\end{itemize}
\end{Thm}

\subsection{Proof of (i)}

The functional 
\[
 \langle \Psi, [H_m(P)-\frac{1}{2\Nm}P^2]\Psi \rangle \ ,
\]
is linear in $P$ and hence $E_m(P)-\frac{1}{2\Nm}P^2$, being the infimum of this expression over all normalized vectors $\Psi$, is a concave function of $P$. Thus $f(P)$ is convex.  $\Box$

\subsection{Proof of (ii)}

Let $T$ be  the time reversal operator  which is defined by
complex conjugating the wave function, reversing all photon momenta,
multiplying by $(-1)^{\one\otimes N_2}$ where $N_2:=\dG(0\oplus\one)$ is the number operator of photons in the $2$
polarization state
and multiplying the spinor by $\Pi_{j=1}^N\sigma_{j2}$ with
$\sigma_j=(\sigma_{j1},\sigma_{j2}, \sigma_{j3})\ j=1,\dots,N$. Clearly $TP_{\mathrm{f}}T =
-P_{\mathrm{f}}$, $TA(x_j)T=-A(x_j)$ and $TB(x_j)T=-B(x_j)$. Moreover , $ T \sigma_j T = - \sigma_j$.
Hence $H_m(P)$ and $H_m(-P)$ are (antiunitarily) equivalent and therefore
$E_m(-P)=E_m(P)$. From this the function $f$ introduced in (i)
satisfies $f(-P)=f(P)$. Since  $f$ is convex by (i),
\begin{align*} 
-E_m(0)=f(0)=f\big(\frac{1}{2}P-\frac{1}{2}P\big)\le \frac{1}{2}f(P)+\frac{1}{2}f(-P)=f(P).
\end{align*}
Thus we conclude (ii).  $\Box$

\subsection{Proof of (iii)}

Property (iii) is a direct consequence  of the following general proposition:

\begin{Prop}\label{Loss}
Let $F(P)$ be a function that satisfies the following conditions:
\begin{itemize}
\item[{\rm (a)}] 
$\displaystyle
F(0)\le F(P), 
$
\item[{\rm (b)}]
$\displaystyle
F(P)\le \frac{P^2}{2}+F(0),
$
\item[{\rm (c)}]
$\displaystyle
g(P)=\frac{P^2}{2}-F(P)
$
is a convex function.
\end{itemize}
Then 
\[
 F(P-k)-F(P)\ge \begin{cases}
-|k||P|+\frac{k^2}{2} &\mbox{if $|k|\le |P|$,}\\
-\frac{P^2}{2} & \mbox{if $|k|\ge |P|$ }
\end{cases}.
\]
\end{Prop}
{\it Proof.}\ \ See Appendix A. $\Box$

\subsection{Proof of (iv)}
The inequality $E_{N,\Mass}\le E_{\Mass}(0)$ is a consequence
of the fact that $E_{N,\Mass}$ is given by a less restrictive
minimization problem than $E_{\Mass}(0)$.
To prove the converse we simply note that due to the direct integral
representation of $H_{N,m}$ in terms of   $H_m(P)$ we get that
\begin{equation}
\langle \Psi, H_{N,m} \Psi \rangle \ge
\int_{\BbbR^3} |f(P)|^2 E_m(P) d P
\end{equation}
for some function $f(P)$ with $\int_{\BbbR^3} |f(P)|^2 d P=1$.
Since, by assumption $E_m(0) \le E_m(P)$, the claim is proved.
$\Box$

\section{Existence of the ground state  for   massive photons}

In this section, we concentrate  on the existence of a
ground state with massive photons, $\Mass>0$.
Throughout this section, we assume  (V.1), (V.2), (E.I.) and $\Mass>0$.

Let $\Sigma_m(P)$ be the threshold energy $\Sigma(P)$ in the case of  massive
 photons.
Likewise let  $\Lambda_m$ be the set of $P$'s satisfying the binding
 condition for the massive case.

\begin{Thm}\label{Existence2}
Assume that $\Lambda_m\neq\emptyset$. Then,
for  $P\in\Lambda_m$ and $|P|< \Nm$,   $H_m(P)$ has
 a ground state.
\end{Thm}

 We will prove this theorem by series of propositions and lemmata.
The basic idea of our proof is taken from  \cite{LLG}.
 The easiest case $N=1$ will be worked and explicitely. It is not hard to extend this proof
to general $N$.

First we prove the following.
\begin{Prop}\label{Existence1}
Let $\Delta_{\Mass}(P)=\inf_k[E_{\Mass}(P-k)-E_{\Mass}(P)+\omega_{\Mass}(k)]$
 and 
let $\delta_m(P)=\min\{\Delta_m(P), \Sigma_m(P)-E_m(P)\}$.
For all $P\in\BbbR^3$, 
\[
 \inf\mathrm{ess.\, spec}(H_m(P))\ge E_m(P)+\delta_m(P).
\]
\end{Prop}

We first  need some preprations.
Let $j_1$ and $j_2$ be two smooth localization functions so that
$j_1^2+j_2^2=1$ and $j_1$ is supported in a ball of radius $L$.
We introduce linear operators $\tilde{j}_1$ and $\tilde{j}_2$ on
$\oplus^{2}L^2(\BbbR^3)$ by
\[
\tilde{j}_i(f_1 \oplus f_{2})=j_i(-\im\nabla_k)f_1\oplus j_i(-\im\nabla_k)
 f_{2}, \ \ i=1,2.
\]
Now we define $j:\oplus^{2}L^2(\BbbR^3)\to
[\oplus^{2}L^2(\BbbR^3)]\oplus [\oplus^{2}L^2(\BbbR^3)]$ by
$jf=\tilde{j_1}f\oplus\tilde{j}_2f$ for each $f\in
\oplus^{2}L^2(\BbbR^3)$. Note that  $j^*j=\one$.

Let $\mathsf{U}$ be the isometry from $\Fock([\oplus^{2}L^2(\BbbR^3)]\oplus
[\oplus^{2}L^2(\BbbR^3)])$ to $ \mathcal{F}\otimes \mathcal{F}$ given in
Section \ref{SecondQ} and set
\[
 \check{\Gamma}(j)=\mathsf{U}\Gamma(j): \mathcal{F}\to \mathcal{F}\otimes \mathcal{F}.
\] 
From the definition  it follows that 
\[
 \check{\Gamma}(j)a_r(f)^{\#}=[a_r(j_1(-\im \nabla_k)f)^{\#}\otimes\one +\one \otimes
 a_r(j_2(-\im\nabla_k)f)^{\#}]\check{\Gamma}(j).
\]
 Since $j$ is an  isometry, so is $\check{\Gamma}(j)$. We remark that, for a multiplication
 operator $h$ on $L^2(\BbbR^3)$, 
\begin{align}
&\big\|\big\{\dG(\oplus^2
 h)-\check{\Gamma}(j)^*[\dG(\oplus^2h)\otimes\one+\one\otimes\dG(\oplus^2h)]
\check{\Gamma}(j)\big\}\Psi\big\|\nonumber\\
\le& \big(\big\|[j_1(-\im\nabla_k),h\big]\big\|+\big\|\big[j_1(-\im\nabla_k),h\big]\big\|\big)\|\Num\Psi\|\label{EstNum}
\end{align}
holds by the definition (or see, e.g., \cite[Section 2]{DG1}).

Let $H^{\otimes}_m(P)$ be a self-adjoint operator on $\Hil^N\otimes
\mathcal{F}\ (N=1)$ assoiated with  the
form sum
\begin{align}
&\frac{1}{2\emass}\Big\{\sigma\cdot\Big(p\otimes\one\ +e\one\otimes
 A(x_1)\Big)\Big\}^2\otimes\one\nonumber\\
+&\frac{1}{2\Nm}\Big(P-p\otimes\one\otimes\one-\one\otimes
 P_{\mathrm{f}}\otimes\one-\one\otimes
\one \otimes P_{\mathrm{f}}-Ze\one\otimes A(0)\otimes \one \Big)^2\nonumber\\
+&\tilde{V}\otimes\one\otimes\one
+\one\otimes H_{\mathrm{f},\Mass}\otimes\one+\one\otimes\one\otimes H_{\mathrm{f},\Mass},\label{Longterm}
\end{align}
where $p=-\im \nabla_{x_1}$. Note that $H^{\otimes}_m(P)$ can be written as 
\[
 H_{1,m}\otimes \one +\one \otimes H_{\mathrm{f},m}+J^{\otimes}(P),
\]
where $J^{\otimes}(P)$ is defined by the second term in (\ref{Longterm}).

\begin{lemm}\label{Iroiro} {\rm (i)}  For $\vphi\in\Hilfin^N$, \[				    
 \la\vphi,
 H_m(P)\vphi\ra=\la\check{\Gamma}(j)\vphi,H^{\otimes}_m(P)\check{\Gamma}(j)\vphi\ra+o_L(\vphi)
\]
where $o_L(\vphi)$ is the error term which satisfies 
\[
 |o_L(\vphi)|\le \tilde{o}(L^0)(\|H_m(P)\vphi\|^2+\|\vphi\|^2).
\]
Here $\tilde{o}(L^0)$ is a function of  $L$ does not depend on $\vphi$
		  and vanishes as $L\to \infty$.

 {\rm (ii)} For $\vphi\in\Hil_{\mathrm{fin}}^N\hat{\otimes}\mathfrak{F}_{\mathrm{fin}}(\oplus^2C_0^{\infty}(\BbbR^3))$,
\begin{align*} 
\la\vphi, H^{\otimes}_m(P)\vphi\ra \ge\la\vphi,[ E_m(P)+(\one-P_{\Omega})\Delta_m(P)]\vphi\ra,
\end{align*}
where $P_{\Omega}$ is the orthogonal projection onto $\Hil^N\otimes\Omega$.
\end{lemm}
$Proof.$\ \  (i) In  \cite[Lemma A.1]{LLG}  the following assertion  has
 already been  proven,
\[
\la\vphi, H_{1,m}\vphi\ra=\la\vphi,\check{\Gamma}(j)^*[H_{1,m}\otimes\one+\one\otimes H_{\mathrm{f},m}]\check{\Gamma}(j)\vphi\ra+o_L(\vphi).
\]
So it suffices to prove 
\[
 \la\vphi,J(P)\vphi\ra=\la\vphi,\check{\Gamma}(j)^*J^{\otimes}(P)\check{\Gamma}(j)\vphi\ra+o_L(\vphi),
\]
where 
\[
 J(P)=\frac{1}{2\Nm}\Big(P-p\otimes\one-\one\otimes
 P_{\mathrm{f}}-Ze\one\otimes A(0) \Big)^2.
\]
Let $X=P-p\otimes\one-ZeA(0)$. We can easily check 
\begin{align*}
J(P)-\check{\Gamma}(j)^*J^{\otimes}(P)\check{\Gamma}(j)
=(X-\one\otimes P_{\mathrm{f}})Q+Q(X-\one\otimes P_{\mathrm{f}})-Q^2,
\end{align*}
where
\begin{align*}
 Q=X-\one\otimes
 P_{\mathrm{f}}-\check{\Gamma}(j)^*\Big(X\otimes\one-\one\otimes
 P_{\mathrm{f}}\otimes \one-\one\otimes\one\otimes
 P_{\mathrm{f}}\Big)\check{\Gamma}(j).
\end{align*}
Therefore it is enough to show $\|Q\Psi\|=o_L(\Psi)$ for $\Psi\in\Hilfin^N$.
On the one hand, in \cite[Lemma A. 1]{LLG}, it is  already proven that
\[
 \Big\|\Big(X-\check{\Gamma}(j)^*X\otimes\one\check{\Gamma}(j)\Big)\Psi\Big\|=o_L(\Psi).
\] 
On the other hand, by Theorem \ref{SA},
we have 
\[
 \|\one\otimes N_{\mathrm{f}}\Psi\|\le C\Big(\|H_m(P)\Psi\|+\|\Psi\|\Big)
\]
for some $C>0$ (which depends on $m$) and therefore, by (\ref{EstNum}),
\begin{align*}
&\Big\|\Big[\one\otimes
 P_{\mathrm{f}i}-\check{\Gamma}(j)^*\Big(\one\otimes
 P_{\mathrm{f}i}\otimes \one+\one\otimes\one\otimes
 P_{\mathrm{f}i}\Big)\check{\Gamma}(j)\Big]\Psi\Big\|\\
\le & \big(\big\|[j_1(-\im\nabla_k),k_i\big]\big\|+\big\|\big[j_2(-\im\nabla_k),k_i\big]\big\|\big)\, \|\one\otimes N_{\mathrm{f}}\Psi\|\\
\le &\frac{\mathrm{const.}}{L}\Big(\|H_m(P)\Psi\|+\|\Psi\|\Big),\ \ i=1,2,3,
\end{align*}
where we use the fact $\|[j_l(-\im\nabla_k),k_i]\|\le \mathrm{const.}/L\ (l=1,2)$.
Hence we  have the desired assertion.

(ii)\ \ Before we start the proof, we need some  preparations.
Let $\mathsf{S}_n$ be the permutation group of degree $n$. For
$k^{(n)}=(k_1^{(n)},\dots,k_n^{(n)})\in\BbbR^{3n},\,
k_j^{(n)}\in\BbbR^3$
 and $\sigma\in\mathsf{S}_n$, 
 set $k^{(n)}_{\sigma}=(k_{\sigma(1)}^{(n)},\dots,k_{\sigma(n)}^{(n)})$.
We introduce a closed subspace 
$L^2_{\mathrm{sym}}(\BbbR^{3\alpha_1}\times\BbbR^{3\alpha_2})$
of
$L^2(\BbbR^{3\alpha_1}\times\BbbR^{3\alpha_2})$, 
consists of functions satisfying
\[
 \psi\big(k_{1,\sigma_1}^{(\alpha_1)};k_{2,\sigma_2}^{(\alpha_2)}\big)=
\psi\big(k_1^{(\alpha_1)};k_2^{(\alpha_2)}\big)
\]
for any $\sigma_j\in\mathsf{S}_{\alpha_j},\ j=1,2$.
Let $h$ be a multiplication operator on $L^2(\BbbR^3)$ by the function
$h(k)$.
For $\alpha=(\alpha_1,\alpha_2)\in\BbbN_0^2$,  we define a linear
operator $h^{(\alpha)}$ on
$L^2_{\mathrm{sym}}(\BbbR^{3\alpha_1}\times\BbbR^{3\alpha_2})$
by
\[
 \big(h^{(\alpha)}\psi\big)\big(k_1^{(\alpha_1)};k_2^{(\alpha_{2})}\big)
=\sum_{r=1,2}\sum_{l=1}^{\alpha_r}h\big(k_{rl}^{(\alpha_r)}\big)
\psi\big(k_1^{(\alpha_1)};k_2^{(\alpha_{2})}\big),
\]
where $k_{rl}^{(\alpha_r)}$ is the $l$-th component of
$k_r^{(\alpha_r)}=(k_{r1}^{(\alpha_r)},\dots,k_{r\alpha_r}^{(\alpha_r)})$.
It is well-known that there is a natural identification such that
\begin{align*}
 \mathcal{F}&=\bigoplus_{\alpha_1,\alpha_{2}=0}^{\infty}
L^2_{\mathrm{sym}}(\BbbR^{3\alpha_1}\times\BbbR^{3\alpha_{2}}),\label{UseId}
\ \ \ \dG(\oplus^2 h)=\bigoplus_{\alpha_1,\alpha_{2}=0}^{\infty}h^{(\alpha)}.
\end{align*}

 Note that the Hilbert space $\Hil^N\otimes\mathcal{F}$ has the following direct sum
decomposition:
\[
 \Hil^N\otimes\mathcal{F}=\bigoplus_{\alpha\in\BbbN_0^{2}}\Hil^N\otimes L^2_{\mathrm{sym}}(\BbbR^{3\alpha_1}\times\BbbR^{3\alpha_{2}}).
\]
The restriction of $H_m^{\otimes}(P)$ to the subspace $\Hil^N\otimes
L^2_{\mathrm{sym}}(\BbbR^{3\alpha_1}\times\BbbR^{3\alpha_{2}})$,
 $\alpha\ne (0,0)$,
is given by
\begin{align*}
(H_m^{\otimes}\Psi)(k_1^{(\alpha_1)};k_2^{(\alpha_2)})=&H_m\Big(P-\sum_{r=1,2}\sum_{l=1}^{\alpha_r}
 k_{rl}^{(\alpha_r)}\Big)\Psi(k_1^{(\alpha_1)};k_2^{(\alpha_2)})\\
&+\sum_{r=1,2}\sum_{l=1}^{\alpha_r} \omega_{\Mass}(k_{rl}^{(\alpha_r)})\Psi(k_1^{(\alpha_1)};k_2^{(\alpha_2)})
\end{align*}
for $\Psi\in \Hil^N\otimes
L^2_{\mathrm{sym}}(\BbbR^{3\alpha_1}\times\BbbR^{3\alpha_{2}})$. Thus 
\begin{align}
\la\Psi, H^{\otimes}_m(P)\Psi\ra\ge&\int\Big[E_m\Big(P-\sum_{r=1,2}\sum_{l=1}^{\alpha_r}
 k_{rl}^{(\alpha_r)}\Big)+\sum_{r=1,2}\sum_{l=1}^{\alpha_r}
 \omega_{\Mass}(k_{rl}^{(\alpha_r)})\Big]\nonumber\\
&\times\big\|\Psi(k_1^{(\alpha_1)};k_2^{(\alpha_2)})\big\|^2_{L^2(\BbbR^3)\otimes\mathcal{F}}\,
 \mathrm{d}k_1^{(\alpha_1)}\mathrm{d}k_2^{(\alpha_2)}\nonumber\\
&\ge\big(\Delta_m(P)+E_m(P)\big)\|\Psi\|^2,\label{DelPE}
\end{align}
where we use the fact $\omega_m(k_1)+\omega_m(k_2)\ge \omega_m(k_1+k_2)$.
On the other hand, on ``0-particle space'' $\Hil^N\otimes\Omega$, we have
\begin{align}
\la\vphi\otimes\Omega, H^{\otimes}_m(P)\vphi\otimes\Omega\ra\ge
 E_m(P)\|\vphi\otimes\Omega\|^2.
\label{DelEP2}
\end{align}
Combining  (\ref{DelPE}) and (\ref{DelEP2})  we obtain (ii). $\Box$
\vspace{0.5cm}

Let $\phi$ and $\bar{\phi}$ be  nonnegative $C^{\infty}$  functions with
$\phi^2+\bar{\phi}^2=1$, $\phi$ identically $1$ on the unit ball, and 
vanishing outside the ball of radius $2$. Let $\phi_R(x)=\phi(x/R)$.
For any $\Psi\in\Hilfin^N$,
\begin{align}
\la\Psi,
 H_m(P)\Psi\ra=&\la\phi_R\Psi,H_m(P)\phi_R\Psi\ra+\la\bar{\phi}_R\Psi,H_m(P)\bar{\phi}_R\Psi\ra\nonumber\\
&-C\la\Psi, (\nabla\phi_R)^2\Psi\ra-C\la\Psi, (\nabla\bar{\phi}_R)^2\Psi\ra\label{Rapp}
\end{align}
with $C=1/2\Nm+1/2\emass$.
The last two term vanish,  if we take $R\to \infty$.  
 
Let
\[
 \Sigma_{m,R}(P)=\inf_{\vphi\in\mathcal{D}_R,\,  \|\vphi\|=1}\la\vphi,H_m(P)\vphi\ra.
\]

\begin{lemm}\label{EstBelow}
For all $\Psi\in\D(H_m(P))$  we have
\begin{align}
&\la\Psi, H_m(P)\Psi\ra\nonumber\\
\ge&(E_m(P)+\delta_{m,R}(P))\|\Psi\|^2-\Delta_m(P)\|\phi_R\Gamma(\tilde{j}_1)\Psi\|^2+o(1)\|\Psi\|_{H_m(P)}^2, \label{LowBound}
\end{align}
where $\delta_{m,R}(P)=\min\{\Delta_m(P),\Sigma_{m,R}(P)-E_m(P)\}$, $o(1)$
 is the error term  vanishing uniformly in $\Psi$  as both $L, R\to
 \infty$ and $\|\Psi\|^2_{H_m(P)}:=\|H_{m}(P)\Psi\|^2+\|\Psi\|^2$.
\end{lemm}
{\it Proof.}\ \ Clearly
\begin{align*}
\la\bar{\phi}_R\Psi,H_m(P)\bar{\phi}_R\Psi\ra\ge \Sigma_{m,R}(P)\|\bar{\phi}_R\Psi\|^2.
\end{align*}
Thus, noting
$\|P_{\Omega}\check{\Gamma}(j)\Phi\|=\|\Gamma(\tilde{j}_1)\Phi\|$, 
 we  obtain (\ref{LowBound}) by Lemma \ref{Iroiro} and (\ref{Rapp}) for
 $\Psi\in\Hilfin^N$. Since $\Hilfin^N$ is a core of $H_m(P)$, 
  this inequality extends  to $\D(H_m(P))$. 
 $\Box$ \medskip\\
{\sf Proof of Proposition \ref{Existence1}}\\
For any $\lambda\in\mathrm{ess.\, spec}(H_m(P))$, there is a sequence
$\{\Psi_n\}$ such that $\|\Psi_n\|=1,\ \wlim\Psi_n=0,$ and 
$\lim_{n\to\infty}\|(H_m(P)-\lambda)\Psi_n\|=0$.
For any $n\in\BbbN$, 
\[
 \la\Psi_n,
 H_m(P)\Psi_n\ra\ge E_m(P)+\delta_{m,R}(P)-\Delta_m(P)\|\phi_R\Gamma(\tilde{j}_1)\Psi_n\|^2+o(1)\|\Psi_n\|^2_{H_m(P)}
\]
by Lemma \ref{EstBelow}.
First, take $n\to \infty$. Notice that
\[
\|\phi_R\Gamma(\tilde{j}_1)\Psi_n\|^2=\la\phi_R^2\Gamma(\tilde{j}_1^2)\Psi_n,
(\one+p^2\otimes\one+\one\otimes
H_{\mathrm{f},\Mass})^{-1/2}(\one+p^2\otimes \one+\one\otimes H_{\mathrm{f},\Mass})^{1/2}\Psi_n\ra. 
\]
Since $(\one+p^2\otimes\one+\one\otimes
H_{\mathrm{f},\Mass})^{-1/2}\phi_R\Gamma(\tilde{j}_1)$ is compact on every finite
particle space and $\la\Psi_n, \one\otimes N_{\mathrm{f}}\Psi_n\ra$ is uniformly bounded
on account of the positive photon mass, we have $\|\phi_R\Gamma(\tilde{j}_1)\Psi_n\|\to 0$ as $n\to
\infty$ and $\lambda\ge E_m(P)+\delta_{m,R}(P)+o(1)(\lambda^2+1)$.
Taking $R\to\infty$  and $L\to \infty$, we obtain $\lambda\ge
E_m(P)+\delta_m(P)$.
This means 
\[
 \inf \mathrm{ess.\, spec}(H_m(P))\ge E_m(P)+\delta_{m}(P).\ \ \Box
\]

\begin{Prop}\label{Positve}
For $|P|< \Nm$, $\Delta_m(P)>0$.
\end{Prop}
{\it Proof.}\ \
Let $\Delta_m(P:k)=E_m(P-k)-E_m(P)+\omega_m(k)$.
Note that $\Delta_m(P)\ge \min\{\inf_{|k|\le|P|}\Delta_m(P:k), \inf_{|k|\ge
|P|}\Delta_m(P:k)\}$. Thus, it sufficies to show that
$\inf_{|k|\le|P|}\Delta_m(P:k)>0$ and $ \inf_{|k|\ge|P|}\Delta_m(P:k)>0$.
Applying Theorem \ref{IneqGSE1} (iii), we obtain
\[
 \inf_{|k|\le|P|}\Delta_m(P:k)\ge \inf_{|k|\le|P|}\Big\{\omega_{\Mass}(k)-\frac{|k||P|}{\Nm}\Big\}>0
\]
whenever $|P|<\Nm$. Moreover, again by Theorem \ref{IneqGSE1} (iii), 
\begin{align*}
\inf_{|k|\ge|P|}\Delta_m(P:k)\ge \inf_{|k|\ge
 |P|}\Big\{-\frac{P^2}{2\Nm}+\omega_{\Mass}(k)\Big\}\\
=-\frac{P^2}{2\Nm}+\sqrt{P^2+\Mass^2}>0 
\end{align*}
whenever
$|P|<\sqrt{2\Nm\Big(\Nm+\sqrt{\Nm^2+\Mass^2}\Big)}$. $\Box$\medskip\\
{\sf Proof of Theorem \ref{Existence2}}\\
By Proposition \ref{Positve}, $\delta_m(P)>0$ for $P\in\Lambda_m$.
Thus, by Proposition \ref{Existence1}, one has  $\inf \mathrm{ess.\,
spec}(H_m(P))-E_m(P)\ge \delta_m(P)>0$, which implies Theorem
\ref{Existence2}. $\Box$

\section{Proof of Theorems \ref{ExMassless}, \ref{MainTheorem}, and \ref{Iontype}}

\subsection{Exponential decay}

By the following lemma we can reduce the binding condition with   massive
photons   to the one with  massless ones. 

\begin{lemm}\label{MasslesstoMass}
\begin{itemize}
\item[{\rm (i)}] $E_{\Mass}(P)\to E(P)$ as $\Mass\to 0$.
\item[{\rm (ii)}] $\Sigma_{\Mass}(P)$ is a convergent sequence and 
$\displaystyle \lim_{\Mass\to 0}\Sigma_{\Mass}(P)\ge \Sigma(P)$.
\item[{\rm (iii)}] Suppose that $P\in\Lambda$. Then there exists $\mm>0$
		 such that, for all $\mm>m\ge0$,  $P\in \Lambda_m$.
\end{itemize}
\end{lemm}
{\it Proof.}\ \ (i) For $m_1\ge m_2$, $H_{m_1}(P)\ge H_{m_2}(P)$. Thus
$\{E_m(P)\}$ is monotonically decreasing and $\lim_{m\to 0}E_m(P)$
exists. Clearly $E(P)\le \lim_{m\to0}E_m(P)$.

We will prove $E(P)\ge \lim_{m\to 0}E_m(P)$.
For arbitrary $\vepsilon>0$, there is $\vphi\in\Hil_{\mathrm{fin}}^N$
such that $\|\vphi\|=1$ and 
\[
 \la\vphi, H(P)\vphi\ra\le E(P)+\vepsilon.
\]
Noting  $H_m(P)\le H(P)+m\one\otimes N_{\mathrm{f}}$,
we have
\begin{align*}
E_m(P)& \le \la \vphi, H_m(P)\vphi\ra\\
&\le \la \vphi, [H(P)+m\one\otimes N_{\mathrm{f}}]\vphi\ra\\
&\le E(P)+\vepsilon+m\|\one\otimes N_{\mathrm{f}}^{1/2}\vphi\|^2.
\end{align*}
Taking the limit $m\to 0$  we obtain 
\[
 \lim_{m\to 0}E_m(P)\le E(P)+\vepsilon.
\]
Since $\vepsilon>0$ is arbitrary, $\lim_{m\to 0}E_m(P)\le E(P)$ follows.

(ii) For $m_1\ge m_2$, we can easily see that $\Sigma_{m_1}(P)\ge
\Sigma_{m_2}(P)$. Accordingly, $\{\Sigma_{\Mass}(P)\}$ is  monotically
decreasing and has a finite limit $\tilde{\Sigma}(P):=\lim_{m\to
0}\Sigma_m(P)$. Note that, for all $m>0$, $\Sigma_m(P)\ge \Sigma(P)$.
Thus we have $\tilde{\Sigma}(P)\ge \Sigma(P)$.

(iii) Let $P\in\Lambda$. Then $\alpha=\Sigma(P)-E(P)>0$.
For all $\vepsilon>0$ so that $\alpha-2\vepsilon>0$,
there is a $\mm>0$ so that, for all $m$ with $\mm-m<\vepsilon$,
$|\tilde{\Sigma}(P)-\Sigma_m(P)|<\vepsilon$ and
$|E(P)-E_m(P)|<\vepsilon$.
Then
\begin{align*}
\Sigma_m(P)-E_m(P)&=\Sigma(P)-E(P)+\big\{(\Sigma_m(P)-\tilde{\Sigma}(P))+(\tilde{\Sigma}(P)-\Sigma(P))\\
&\ \ \ +(E(P)-E_m(P))\big\}\\
&\ge \alpha -2\vepsilon>0.
\end{align*}
This means $P\in\Lambda_m$ if $m<\mm$. $\Box$

\begin{lemm}\label{ExDecay}
Let  $\beta$ be a real numbers and $\alpha=\sum_{j=1}^N(1/2\emass)+N/\Nm$.
 For $P\in\Lambda$  suppose that
 $E(P)+\alpha\beta^2<\Sigma(P)$. For each $P\in\Lambda$ and $|P|<\Nm$, let $\Gr$ be a
 normalized ground state for $H_{\Mass}(P)$. Then, for $m>0$ sufficiently small and
 $R$ sufficiently large,
\[
 \big\|\ex^{\beta|x|}\Psi_{P,m}\big\|^2\le
 C_{\beta}\ \ex^{4\beta R}\Big(\frac{1}{\Sigma(P)-E(P)-\alpha\beta^2+o(1)}+1\Big),
\]
where $C_{\beta}$ is a positive constant depends on  $\beta$ but
 independent of $R, \Mass$, and $o(1)$ is
 the error term  vanishing as $m\to 0$ and $R\to \infty$.
\end{lemm}
\begin{rem}
{\rm
Existence of $\Psi_{P,m}$ is guaranteed by Theorem \ref{Existence2} and 
Lemma \ref{MasslesstoMass} for small $m$.
}
\end{rem}
{\it Proof.}\ \   Note first that each $G\in C^{\infty}(\BbbR^{3N})$
with $G,|\nabla_j G|\in L^{\infty}(\BbbR^{3N})$,
\begin{align}
 [[H_m(P),G],G]=-\sum_{j=1}^{N}\frac{1}{\emass}\big(\nabla_j
 G\big)^2-\frac{1}{\Nm}\Big(\sum_{j=1}^{N}\nabla_j G\Big)^2.\label{ExDc}
\end{align}
Take $G(x)=\chi(x/R)\ex^{f(x)}$ where $f(x)=\beta|x|/(1+\varepsilon |x|)$ and
$0\le \chi\le 1$ 
is a smooth function 
that is identically $1$ outside the ball radius $2$, and $0$ inside the
ball radius $1$. 
With a slight modification of \cite[Proof of Lemma 6.2]{LLG}, we
get
\begin{align}
\Big\la
&G\Psi_{P,m},\Big\{H_m(P)-E_m(P)-\sum_{j=1}^N\frac{1}{2\emass}|\nabla_j
f|^2-\frac{1}{2m_{\mathrm{n}}}\Big(\sum_{j=1}^N\nabla_jf\Big)^2\Big\}G\Psi_{P,m}\Big\ra\nonumber\\
\le& C_{\beta}\, \ex^{4\beta R}\label{ExDc2}
\end{align}
by (\ref{ExDc}). Using the facts $\Sigma_{m,R}(P)\ge \Sigma_{0,R}(P)$ for
any $m$, $|\nabla_j f|\le \beta$, and Lemma \ref{MasslesstoMass} (i), we obtain 
\begin{align*}
\mbox{LHS of (\ref{ExDc2})}&\ge
 (\Sigma_{m,R}(P)-E_m(P)-\alpha\beta^2)\|G\Psi_{P,m}\|^2\\
&\ge \big\{\Sigma(P)-E(P)-\alpha\beta^2+(\Sigma_{0,R}(P)-\Sigma(P))\\&\hspace{0.5cm}+(E(P)-E_m(P))\big\}\|G\Psi_{p,m}\|^2\\
&=\big(\Sigma(P)-E(P)-\alpha\beta^2+o(1)\big)\|G\Psi_{P,m}\|^2.
\end{align*}
Therefore the assertion follows by taking $\varepsilon \to 0$. $\Box$

\subsection{A photon number  bound and photon derivative bound}

Let
\begin{align*}
\mathfrak{P}_j&=-\im\nabla_j\otimes\one+e A(x_j),\ \ j=1,\dots,N,\\
\mathfrak{P}_0&=P+\im\sum_{j=1}^{N}\nabla_j\otimes\one-\one\otimes
 P_{\mathrm{f}}-Ze A(0).
\end{align*}

For later use we   first prove the following. 
\begin{lemm}\label{ResBound} Assume (V.1), (V.2) and (E.I.).
Suppose  that $|P|<\Nm$. Let $\Delta_{\Mass}(P:k):=E_{\Mass}(P-k)-E_{\Mass}(P)+\omega_{\Mass}(k)$.
Then the following assertion  hold for any $\Mass\ge 0$, coupling $e$,
 and cutoffs $\sigma,\kappa$.
\begin{itemize}
\item[{\rm (i)}] $\displaystyle \Delta_{\Mass}(P:k)\ge (1-|P|/\Nm)|k|$. Thus
		  $H_{\Mass}(P-k)-E_{\Mass}(P)+\omega_{\Mass}(k)$
                 has the bounded inverse, denoted by  $\Res$, for $k\neq 0$.
\item[{\rm (ii)}] $\displaystyle \|\Res\|\le C/|k|$, where $C$ is a
		 positive constant independent of $\Mass$ and $k$.
\item[{\rm (iii)}] $\displaystyle \| \mathfrak{P}_{j,l}\Res\|\le
		 C(1+1/|k|),\  j=0,1,\dots,N,\ l=1,2,3$.
\item[{\rm (iv)}] $\displaystyle \|
		 [H_{\Mass}(P)-E_{\Mass}(P)]\Res\|\chi_{0,\kappa}(k) \le  C(1+\kappa)\chi_{0,\kappa}(k).$ 
\end{itemize}
\end{lemm}
{\it Proof.}\ \ (i) If $|k|\le |P|$, the claim follows  by Theorem
\ref{IneqGSE1} (iii).  Suppose that $|k|>|P|$. Then, since
$|P||k|/\Nm>P^2/2\Nm$, we have
\begin{align*}
\Delta_{\Mass}(P:k)&\ge E_{\Mass}(P-k)-E_{\Mass}(P)+|k|\\
&\ge |k|-\frac{P^2}{2\Nm}
\ge |k|-\frac{|P||k|}{\Nm}
=\Big(1-\frac{|P|}{\Nm}\Big)|k|
\end{align*}
by Theorem \ref{IneqGSE1}.  (ii)  immediately follows from (i).

(iii) This is a direct consequence of Lemma \ref{UniEst} and (i).

(iv) Note that 
\[
 [H_{\Mass}(P)-E_{\Mass}(P)]\Res=\one +[H_{\Mass}(P)-H_{\Mass}(P-k)-\omega_{\Mass}(k)]\Res.
\]
For all $\Psi\in\Hil^{N}_{\mathrm{fin}},$
\[
 [H_{\Mass}(P)-H_{\Mass}(P-k)-\omega_{\Mass}(k)]\Psi=[2k\cdot(k-\mathfrak{P}_0)-(k^2+\omega_{\Mass}(k))]\Psi.
\]
Therefore
\[
 \big\|\big(H_{\Mass}(P)-H_{\Mass}(P-k)-\omega_{\Mass}(k)\big)\Psi\big\|\le 2|k|\sum_{j=1}^3 \|(\mathfrak{P}_{0,j}-k_j)\Psi\|+(k^2+\omega_{\Mass}(k))\|\Psi\|.
\]
Since there  is a constant $C$ independent of $P, \Mass$ and $k$ such that 
\[
 \|(\mathfrak{P}_{0,j}-k_j)\Psi\|\le
 C\big(\|H_{\Mass}(P-k)\Psi\|+\|\Psi\|\big),\ \ j=1,2,3,
\]
by Lemma \ref{UniEst},
one has 
\begin{align*}
&\big\| [H_{\Mass}(P)-H_{\Mass}(P-k)-\omega_{\Mass}(k)]\Res\Psi\big\|\\
&\le C \Big[|k|\Big(\|
 H_{\Mass}(P-k)\Res \Psi\|+\|\Res\Psi\|\Big)\\
&\ \ +(k^2+\omega_{\Mass}(k))\|\Res\Psi\|\Big].
\end{align*}
Notice that 
\[
  H_{\Mass}(P-k)\Res\Psi=\Big\{\one +\Res [E_{\Mass}(P)-\omega_{\Mass}(k)]\Big\}\Psi.
\]
Thus, considering $\Delta_{\Mass}(P:k)\ge (1-|P|/\Nm)|k|$ by (i),
\begin{align*}
\| H_{\Mass}(P-k)\Res\|&\le 1+\Delta_{\Mass}(P:k)^{-1}|E_{\Mass}(P)-\omega_{\Mass}(k)|\\
&\le C\, (1+|k|^{-1})
\end{align*}
for $|k|\le |P|$.  As for $\omega_{\Mass}(k)\|\Res\|\, (|k|\le |P|)$, we have to
be more careful. By Theorem \ref{IneqGSE1} (iii),
\[
 \|\Res\|\le \Delta_{\Mass}(P:k)^{-1}\le [\omega_{\Mass}(k)-|k||P|/\Nm]^{-1}
\]
and hence 
\begin{align*}
\omega_{\Mass}(k) \|\Res\|&\le \omega_{\Mass}(k)\big[\omega_{\Mass}(k)-|k||P|/\Nm\big]^{-1}\\
&=1+\frac{|k||P|/\Nm}{\omega_{\Mass}(k)-|k||P|/\Nm}\\
&\le 1+\frac{|P|}{\Nm-|P|}<\infty. 
\end{align*}
Combining these results, one concludes that
\[
\big\|[H_{\Mass}(P)-E_{\Mass}(P)]\Res\big\|\chi_{0,\kappa}(k)\le
C\,  \kappa \chi_{0,\kappa}(k) 
\]
for $|k|\le |P|$.

Similarly, we have, for $|k|>|P|$,
\begin{align*}
|k|\, \| H_{\Mass}(P-k)\Res\|&\le C\, |k|,\\
|k|\, \|\Res\|&\le C,\\
\omega_{\Mass}(k)\|\Res\|&\le C. 
\end{align*}
Hence the assertion follows.  $\Box$

\begin{Prop}\label{PhotonNBound}{\rm (photon number bound)}
Assume   (V.1), (V.2), (E.I.)  and (N). Suppose that $\sigma=0$.
 Then
\[
 \|a_r(k)\Gr\|\le C_{\kappa} \frac{\chi_{0,\kappa}(k)}{|k|^{1/2}},
\] 
where $C_{\kappa}$ is a positive constant independent of $k$ and $\Mass$,
 but depends on $\kappa$ .
\end{Prop}
{\it Proof.}\ \
From the pull-through formula for $a_r(k)$  one concludes
\begin{align*}
a_r(k)H_{\Mass}(P)\Psi_{P,\Mass}=&[H_{\Mass}(P-k)+\omega_{\Mass}(k)]a_r(k)\Psi_{P,\Mass}
\\
&-\sum_{j=1}^{N} \frac{1}{\emass}\mathfrak{P}_j\cdot
\mathfrak{K}^{(m)}_{j,r}(x_j,k)\Psi_{P.\Mass}\\
&-\frac{1}{\Nm}\mathfrak{P}_0\cdot
 \mathfrak{K}^{(m)}_{0,r}(0,k)\Psi_{P,\Mass}\\
&-\sum_{j=1}^N\frac{\im\sigma_j}{2\emass}\cdot k\wedge \mathfrak{K}^{(m)}_{j,r}(x_j,k)
 \Psi_{P,\Mass}
-\frac{\im\sigma_0}{2\Nm}\cdot k\wedge\mathfrak{K}^{(m)}_{0,r}(0,k)\Psi_{P,\Mass},
\end{align*}
where
\[
 \mathfrak{K}^{(m)}_{j,r}(k,x):=e_j \frac{\chi_{0,\kappa}(k)e^r(k)}{\sqrt{2(2\pi)^3\omega_m(k)}}\,
 \ex^{-\im k\cdot x},\ \ j=0, 1,\dots,N,\ r=1,2,3
\]
with $e_0=Ze$ and $e_j=-e$ for $j=1,\dots,N$.
(Note that  in the above  we use  $k\cdot e^r(k)=0$.)
Thus it follows that
\begin{align*}
&\big[H_{\Mass}(P-k)-E_{\Mass}(P)+\omega_{\Mass}(k)\big]a_{\lambda}(k)\Gr\\
=&\sum_{j=0}^N\frac{1}{m_j}\mathfrak{P}_j\cdot\mathfrak{K}^{(m)}_{j,r}(0,k)\Gr+\sum_{j=1}^{N}\frac{1}{m_j}\mathfrak{P}_j\cdot
 \delta\mathfrak{K}^{(m)}_{j,r}(x_j,k)\Gr\\
&+\sum_{j=1}^N\frac{\im\sigma_j}{2\emass}\cdot k\wedge \mathfrak{K}^{(m)}_{j,r}(x_j,k)
 \Psi_{P,\Mass}
+\frac{\im\sigma_0}{2\Nm}\cdot k\wedge\mathfrak{K}^{(m)}_{0,r}(0,k)\Psi_{P,\Mass},
\end{align*}
where
$\delta\mathfrak{K}^{(m)}_{j,r}(x,k):=\mathfrak{K}^{(m)}_{j,r}(x,k)-\mathfrak{K}^{(m)}_{j,r}(0,k)$
and $m_0=\Nm,\  m_j=\emass\, (j=1,\dots,N)$.

 By Lemma \ref{ResBound} (i),
$H_{\Mass}(P-k)-E_{\Mass}(P)+\omega_{\Mass}(k)$ has the bounded inverse $\Res$
for any $\Mass\ge 0$. 
We also note that
\[
 \frac{1}{\emass}\mathfrak{P}_j=\im[H_{\Mass}(P),x_j]+\frac{1}{\Nm}\mathfrak{P}_0,\
 \ j=1,\dots,N.
\]
Accordingly we have
\begin{align}
&a_r(k)\Gr\nonumber\\
=&\im\Res[H_{\Mass}(P)-E_{\Mass}(P)]\sum_{j=1}^{N}\mathfrak{K}^{(m)}_{j,r}(0,k)\cdot
 x_j\Gr\nonumber\\
&+\frac{1}{\Nm}\sum_{j=0}^N\Res\mathfrak{P}_0\cdot\mathfrak{K}^{(m)}_{j,r}(0,k)\Gr\nonumber\nonumber\\
&+\sum_{j=1}^{N}\frac{1}{\emass}\Res\mathfrak{P}_j\cdot\delta\mathfrak{K}^{(m)}_{j,r}(x_j,k)\Gr\nonumber\\
&+\sum_{j=1}\Res\frac{\im\sigma_j}{2\emass}\cdot k\wedge \mathfrak{K}^{(m)}_{j,r}(x_j,k)
 \Psi_{P,\Mass}
+\Res\frac{\im\sigma_0}{2\Nm}\cdot
 k\wedge\mathfrak{K}^{(m)}_{0,r}(0,k)\Psi_{P,\Mass}\nonumber\\
=:& I_1(k)+I_2(k)+I_3(k)+I_4(k).\label{aofkGr2}
\end{align}
By the neutrality condition (N),  $I_2=0$ and by Lemma \ref{ExDecay} and
\ref{ResBound},
one concludes that
\begin{align*} 
\|I_1(k)\|,\ \|I_4(k)\|\le \frac{C}{|k|^{1/2}}\chi_{0,\kappa}(k).
\end{align*}
As for $I_3(k)$, noting $|\delta\mathfrak{K}^{(m)}_{j,r}(x_j,k)|\le \{2(2\pi)^3\}^{-1/2}|e_j| |k|^{1/2}|x_j|\chi_{0,\kappa}(k)$,
we have 
\[
 \|I_3(k)\|\le \frac{C_{\kappa}}{|k|^{1/2}}\chi_{0,\kappa}(k)
\]
by Lemma \ref{ResBound}. $\Box$

\begin{lemm}\label{DerRes} Assume
 (V.1), (V.2) and  (E.I.).
For all $\Mass>0$ and $|P|<\Nm$,
\begin{align}
 \nabla_k\Res=\Res\Big[\frac{1}{\Nm}(\mathfrak{P}_0-k)-\frac{k}{\omega_{\Mass}(k)}\Big]\Res\label{DerRes}
\end{align}
in the   operator norm  topology.
\end{lemm}
{\it Proof.}\ \ By the second resolvent formula, we have
\begin{align*}
&\mathcal{R}_{P,\Mass}(k+h)-\Res\\
=&\mathcal{R}_{P,\Mass}(k+h)[H_{\Mass}(P-k)-H_{\Mass}(P-k-h)+\omega_{\Mass}(k)-\omega_{\Mass}(k+h)]\Res\\
=&\mathcal{R}_{P,\Mass}(k+h)\Big[\frac{1}{\Nm}h\cdot(\mathfrak{P}_0-k)-h\cdot\nabla_k\omega_{\Mass}(k)+O(h^2)\Big]\Res.
\end{align*}
Thus passing through the limiting argument,  the assertion (\ref{DerRes})
follows. $\Box$

\begin{Prop}\label{PhotonDerBound}{\rm (photon derivative bound)} Assume
 (V.1), (V.2), (E.I.),  (N) and  $\sigma=0$.
Suppose  that $P\in\Lambda$ and $|P|<\Nm$. Then, for $|k|<\kappa$ and
 $(k_1,k_2)\neq 0$,
\begin{align}
\|\nabla_ka_r(k)\Gr\| \le \frac{C_{\kappa}}{|k|^{1/2}\sqrt{k_1^2+k_2^2}},\label{DerPhot}
\end{align}
where $C_{\kappa}$ is a positive constant independent  of $k, \Mass$.
\end{Prop}
{\it Proof.}\ \ By (\ref{aofkGr2})  we obtain
\begin{align}
&\nabla_k a_r(k)\Gr\nonumber\\
=&\im\big[\nabla_k \Res\big][H_{\Mass}(P)-E_{\Mass}(P)]\sum_{j=1}^{N}\mathfrak{K}^{(m)}_{j,r}(0,k)\cdot
 x_j\Gr\label{First}\\
&+\im\Res[H_{\Mass}(P)-E_{\Mass}(P)]\sum_{j=1}^{N}\nabla_k\big[\mathfrak{K}^{(m)}_{j,r}(0,k)\cdot
 x_j\big]\Gr\label{Second}\\
&+\sum_{j=1}^{N}\frac{1}{\emass}\nabla_k\big[\Res\big]\mathfrak{P}_j\cdot\delta\mathfrak{K}^{(m)}_{j,r}(x_j,k)\Gr\label{Third}\\
&+\sum_{j=1}^{N}\frac{1}{\emass}\Res\nabla_k\big[\mathfrak{P}_j\cdot\delta\mathfrak{K}^{(m)}_{j,r}(x_j,k)\big]\Gr.\label{Fourth}\\
&+\nabla_k I_4(k).
\end{align}
Applying Lemma \ref{ExDecay}, \ref{ResBound} and (\ref{DerRes}), we 
 estimate the norms of (\ref{First}) and (\ref{Second}) to
obtain
\[
 \|(\ref{First})\|,\ \  
 \|(\ref{Third})\|\le \frac{C_{\kappa}}{|k|^{3/2}}.
\]

Considering the fact $|\nabla_k e_r(k)|\le C/\sqrt{k_1^2+k_2^2}
\ (k_1,k_2)\neq (0,0)$, we 
also estimate (\ref{Second}) and (\ref{Fourth}) with results
\[
  \|(\ref{Second})\|,\ \ 
\|(\ref{Fourth})\|\le \frac{C_{\kappa}}{|k|^{1/2}\sqrt{k_1^2+k_2^2}}.
\]
Similarly we can estimate $\|\nabla_k I_4(k)\|$.
This implies the assertion in (\ref{DerPhot}). $\Box$

\subsection{Proof of Theorem \ref{ExMassless}}

This proof is a slight modification of \cite[Theorem 2.1]{LLG} and we
only provide on  the outline, for details, see \cite{LLG}.
For $P\in\Lambda$ and $|P|<\Nm$, $H_m(P)$ has a normalized ground
state $\Gr$ whenever $\Mass$ is sufficiently small by Theorem \ref{Existence2} and
Lemma \ref{MasslesstoMass}.
Take $m_1>m_2>\cdots$ tending to 0 and denote $\Psi_{P,m_j}$ by
$\Psi_{P,j}$.
The sequence $\{\Psi_{P,j}\}$ is a minimizing  sequence for $H(P)$.
Indeed
\[
 E_{m_j}(P)=\la \Psi_{P,j}, H_{m_j}(P)\Psi_{P,j}\ra\ge \la \Psi_{P,j},
 H_0(P)\Psi_{P,j}\ra\ge E(P),
\]
Thus $\la \Psi_{P,j}, H(P)\Psi_{P,j}\ra\to E(P)$ as $j\to\infty$
by Lemma \ref{MasslesstoMass}. Since $\|\Psi_{P,j}\|=1$, there is a
subsequence $\{\Psi_{P,j'}\}$ of $\{\Psi_{P,j}\}$ which has a weak limit
$\Psi_P$.
Because 
\[
 0\le \la\Psi_P,(H(P)-E(P))\Psi_P\ra\le
 \liminf_{j\to\infty}\la\Psi_{P,j'}, (H(P)-E(P))\Psi_{P,j'}\ra=0,
\]
it suffices to  prove that $\|\Psi_P\|=1$. (This means the strong
convergence of $\{\Psi_{P,j'}\}$.)
Note that, by Proposition \ref{PhotonNBound}, 
\[
 \la \Psi_{P,j'},\one\otimes N_{\mathrm{f}}\Psi_{P,j'}\ra\le C<\infty,
\]
where $C$ is a positive constant independent of  $j'$.
Hence it sufficies to show the $L^2$-convergence  of each $n$-photon
component
$\Psi^{(n)}_{P,j'}$, where we write
$\Psi_{P,j'}=\oplus_{n=0}^{\infty}\Psi_{P,j'}^{(n)}$.
From the exponential decay, it follows that, for each $R>0$,
\begin{align*}
 \|\tilde{\chi}_R\Psi_{P,j'}\|&=\|\tilde{\chi}_R\, \ex^{-\beta|x|}\,
 \ex^{\beta|x|}\Psi_{P,j'}\|\\
&\le C\, \ex^{-\beta R},
\end{align*}
where $\tilde{\chi}_R:=1-\chi_R$. Accordingly it suffices to show the
$L^2$-convergence in the  domain $|x|<R$. By Proposition
\ref{PhotonNBound},
$\Psi_{P,j'}^{(n)}(x_1,\dots,x_{N},k_1,\dots,k_n)=0$ if $|k_i|>\kappa$
for some $i$. By putting these facts together, it suffices to show
$L^2$-convergence  for $\Psi_{P,j'}^{(n)}$ restricted to the bounded
domain
\[
 \Omega_R:=\{(x,k_1,\dots,k_n)\, |\, |x|<R,|k_i|<\kappa,
 i=1,\dots,n\}\subset \BbbR^{3(N+n)}.
\]

By Proposition \ref{PhotonDerBound}, $\{\Psi_{P,j'}^{(n)}\}_{j'}$ is a
bounded sequence in $W^{1,p}(\Omega_R)$ for each $p<2$ and $R>0$.
(It is not hard to check that
\[
 \|\nabla_{k_i}\Psi_{P,j'}^{(n)}\|_{L^p(\Omega_R)}^p\le C
\int_{|k|<\kappa}\mathrm{d}k\|\nabla_k
a_r(k)\Psi_{P,j'}\|^p\le Const<\infty
\] 
and
$\|\nabla_x\Psi_{P,j'}^{(n)}\|^p_{L^p(\Omega_R)}\le Const <\infty$.)
From the weak convergence of $\{\Psi_{P,j'}^{(n)}\}$ in $L^2(\Omega_R)$,
$\Psi_{P,j'}^{(n)}$
weakly converges to $\Psi_P^{(n)}$ in $W^{1,p}(\Omega_R)$. Now we can
apply the Rellich-Kondrachov theorem \cite[Theorem 8.9]{LL2}. Then $\{\Psi_{P,j'}^{(n)}\}$
 converges strongly  to $\{\Psi_P^{(n)}\}$ in $L^q(\Omega_R)$ of $1\le q
\le 3p(N+n)/3(N+n)-p$. If we choose $p$ as
$2>p>6(N+n)/[2+3(N+n)]$, we obtain the strong convergence of
$\{\Psi_{P,j'}\}$
in $L^2(\Omega_R)$. $\Box$

\subsection{Proof of Theorem \ref{MainTheorem} and \ref{Iontype}}

Let 
\[
 \Sigma^{(N)}=\lim_{R\to\infty}\Big(\inf_{\vphi\in\tilde{\mathcal{D}}_R,
 \|\vphi\|=1}\la
 \vphi, H_{N}\vphi\ra\Big)
\]
with 
\[
\tilde{\mathcal{D}}_R=\{\vphi\in\Hil_{\mathrm{fin}}^{N+1}\, |\, \vphi(x)=0,\
\mbox{if $|x|<R$}\}.
\]

\begin{lemm}\label{lowerbound}
\begin{itemize}
\item[{\rm (i)}]
For all $P\in\BbbR^3$,
\[
 \Sigma^{(N)}\le \Sigma(P).
\] 
\item[{\rm (ii)}] $\displaystyle \Sigma^{(N)}=\min\big\{E_{\beta}+E_{\bar{\beta}}\, |\,
 \beta\in\Pi_N\, \mathrm{and}\, \beta\neq\emptyset, \{0,1,\dots,N\}\big\}$.
\end{itemize}
\end{lemm}
{\it Proof.}\ \ (i)
Assume that  there is a $P_0\in\BbbR^3$ such that
$\Sigma^{(N)} >\Sigma(P_0)$ and set  $\gamma:=\Sigma^{(N)}-\Sigma(P_0)>0$.
There exists $R_0>0$ so that, for  all $R>R_0$,
$\gamma_R:=\Sigma^{(N)}_{R}-\Sigma_{R}(P_0)>0$. Here $\Sigma^{(N)}_{R}$ and $\Sigma_{R}(P)$
stands for $\inf_{\vphi\in\tilde{\mathcal{D}}_R,\|\vphi\|=1}\la\vphi,
H_{N}\vphi\ra$ and $\inf_{\vphi\in\mathcal{D}_R,\|\vphi\|=1}\la\vphi,
H(P)\vphi\ra$ respectively.
 (Note that $\lim_{R\to\infty}\gamma_R=\gamma$.)
Take $R$ as $R>R_0$. This $R$ is kept fixed in the following.
There is a $\vphi\in\mathcal{D}_R,\ \|\vphi\|=1$  so that
\[
 \la\vphi,H(P_0)\vphi\ra\le \Sigma_{R}^{(N)}-\gamma_R/2.
\]
Since $\la\vphi, H(P)\vphi\ra$ is continuous in $P$, there is a
$\delta>0$ such that, for all $P$ with $|P-P_0|\le \delta$,
\[
 \la\vphi,H(P)\vphi\ra\le \Sigma_{R}^{(N)}-\gamma_R/4.
\] 
Pick $f\in C^{\infty}(\BbbR^3)$ as $\mathrm{supp} f\subseteq \{P\in\BbbR^3\, |\,
|P-P_0|\le \delta\},\ \|f\|=1$ and define $\vphi_f:=f\times
\vphi\in\Hil^N$. Then we have
\[
 \la\vphi_f, UH_{N}U^*\vphi_f\ra\le\Sigma_{R}^{(N)}-\gamma_R/4.
\]
Notice that  $U^*\vphi_f(x)=0$ if $|x|<R/2N$.
Since $\Hil_{\mathrm{fin}}^{N+1}$ is a core of $H_N$, there is a
sequence $\{\vphi_n\}$ in $\Hil_{\mathrm{fin}}^{N+1}$ so that $\|\vphi_n\|=1$,
$\vphi_n\to U^*\vphi_f$ and $H_N\vphi_n\to H_NU^*\vphi_f$ as $n\to
\infty$.
Let $j$ and $\bar{j}$ be $C^{\infty}$ functions with  $j^2+\bar{j}^2=1$,
$j$ identically $1$ on the unit ball and vanishing outside  the ball of
radius $2$.  Set $j_R(x)=j(4Nx/R)$ and $\bar{j}_R(x)=\bar{j}(4Nx/R)$. Then one gets
\begin{align*}
\la \vphi_n,H_N\vphi_n\ra=\la j_R \vphi_n,H_N j_R\vphi_n\ra+\la
 \bar{j}_R \vphi_n,H_N \bar{j}_R\vphi_n\ra+o_R(\vphi_n)
\end{align*}
by the IMS localization formula. 
For all $\vepsilon>0$, there is a $n'$ such that, for all $n>n'$,
\[
 |\la\vphi_n,H_N\vphi_n\ra-\la\vphi_f,UH_NU^*\vphi_f\ra|<\vepsilon.
\]
Thus, for all $n>n'$,
\[
\la j_R\vphi_n, H_Nj_R\vphi_n\ra+ \la \bar{j}_R\vphi_n, H_N\bar{j}_R\vphi_n\ra+o_R(\vphi_n)-\vepsilon\le \Sigma^{(N)}_R-\gamma_R/4.
\]
Since $\bar{j}_R\vphi_n/\|\bar{j}_R\vphi_n\|\in\tilde{\mathcal{D}}_{R/2N}$, we have
\[
\la j_R\vphi_n, H_Nj_R\vphi_n\ra+  \Sigma_{R/2N}^{(N)}\|\bar{j}_R\vphi_n\|^2+o_R(\vphi_n)-\vepsilon\le \Sigma_{R}^{(N)}-\gamma_R/4.
\]
We will discuss the limit $n\to \infty$.
Note that 
\begin{align*}
\la j_R\vphi_n, H_Nj_R\vphi_n\ra&=\la
 j_R^2\vphi_n,H_N\vphi_n\ra+o_R(\vphi_n)\\
&\to \la j_R^2U^*\vphi_f,H_NU^*\vphi_f\ra+o_R(U^*\vphi_f)\ \  (n\to \infty). 
\end{align*}
Here we use the fact $\lim_{n\to\infty}o_R(\vphi_n)=o_R(U^*\vphi_f)$
because \[
	 |o_R(\vphi_n)|\le
	 \tilde{o}(R^0)(\|H_N\vphi_n\|^2+\|\vphi_n\|^2).
\]
By the fact $j_RU^*\vphi_f=0$, we conclude that $\lim_{n\to\infty}\la j_R\vphi_n, H_Nj_R\vphi_n\ra=o_R(U^*\vphi_f)$.
Taking the limit $n\to \infty$, we get 
\[
 \Sigma_{R/2N}^{(N)}\|\bar{j}_RU^*\vphi_f\|^2+o_R(U^*\vphi_f)-\vepsilon\le \Sigma_{R}^{(N)}-\gamma_R/4.
\]
Since $\vepsilon$ is arbitrary and $\bar{j}_RU^*\vphi_f=U^*\vphi_f$, we get
\[
 \Sigma_{R/2N}^{(N)}+o_R(U^*\vphi_f)\le \Sigma_{R}^{(N)}-\gamma_R/4.
\]
Therefore, 
taking the limit $R\to\infty$, we conclude that
\[
 \Sigma^{(N)}\le\Sigma^{(N)}-\gamma/4.
\]
This is a contradiction. Proof of (ii) is a slight modification of the
one of \cite[Theorem 3]{Grie}.   $\Box$
\medskip\\
{\sf Proof of Theorem \ref{MainTheorem}}\\
Note that $\{P\in\BbbR^3\, |\, E(P)\le \Sigma^{(N)}\}\subseteq \Lambda$ by the above
lemma.
By the property $E(P)\le E(0)+P^2/2\Nm$ (Theorem \ref{IneqGSE1}),
one also has
$\{P\in\BbbR^3\, |\, E(0)+P^2/2\Nm\le \Sigma^{(N)}\}\subseteq
\Lambda$.
Considering the facts $E(0)=E_{N}$ (Theorem \ref{IneqGSE1}) and
Lemma \ref{lowerbound} (ii),
we obtain $\{P\in\BbbR^3\, |\, |P|<\sqrt{2\Nm
E_{\mathrm{bin}}}\}\subseteq \Lambda$.
Now Theorem \ref{MainTheorem}  follows from Theorem \ref{ExMassless}.
 $\Box$
\medskip\\
{\sf Proof of Theorem \ref{Iontype}}\\
Basic idea of the proof is almost same as Theorem \ref{ExMassless} and \ref{MainTheorem}.
Since the system  is not neutral, the term $I_2(k)$  in
(\ref{aofkGr2}) does not vanish. We can calculate the contribution of $I_2(k)$
as $|I_2(k)|\le \mathrm{const.}|k|^{-3/2}\chi_{0,\kappa}(k)$ in the photon number bound
and $|\nabla_k I_2(k)|\le \mathrm{const.}|k|^{-3/2}\times
(k_1^2+k_2^2)^{-1/2}$ for $|k|<\kappa$ in the photon
derivative bound by Lemma \ref{ResBound}.
If we take the infrared cutoff $\sigma$ as $\sigma>0$, these
singularities at origin $k=0$ do not influence our proof of Theorem
\ref{ExMassless} and \ref{MainTheorem}, and the same arguments still hold. $\Box$

\section{ Spinless electrons,  Boltzmann statistics }\label{NoSS}

In this section we consider  an arbitrary collection of charges with 
no symmetry condition on the wave function  imposed.
 The Hamiltonian is given by 
\begin{align}
H_{N}=\sum_{j=0}^{N}\frac{1}{2m_j}\Big(-\im \nabla_j\otimes \one
-e_j A(x_j)\Big)^2+V\otimes \one +\one\otimes \Hf. \label{LPFHami}
\end{align}
$H_N$ acts on $[\otimes^{N+1}L^2(\BbbR^3)]\otimes
\mathcal{F}$. 
We require $m_j>0$, while $e_j$ is arbitrary, $j=0,\dots,N$.
Note  that
the neutrality condition (N) can then  be rewritten as 
\begin{align*}
\sum_{j=0}^Ne_j=0. \tag{N'}
\end{align*} 
Moreover,  because we  do not consider any statistics of the particles,
our assumptions for potential are generalized as follows: 
\begin{itemize}
\item[{\rm(V'.1)}] $V$ is a pair potential of the
form
\[
 V(x_0,\dots,x_N)=\sum_{0\le i<j\le N} V_{ij}(x_i-x_j)
\]
 and each
$V_{ij}$ is infinitesimally small with respect to $-\Delta$,  
\item[ {\rm (V'.2)}]
each $V_{ij}$ is in $L^2_{\mathrm{loc}}(\BbbR^3)$ and  $V_{ij}(x)\to 0$ as $|x|\to\infty$.
\end{itemize}

Following the argument  in Section 2.2, $H_N$ admits the decomposition
\[
 H_N=\int^{\oplus}_{\BbbR^3}H(P)\, \mathrm{d}P.
\]

 We estabilish the energy inequality (E.I.).

\begin{Prop}\label{EInq}Assume (V'.1).
Then, the energy inequality (E.I.) holds for arbitrary photon mass
 $m$,  couplings $e_1,\dots, e_N$, 
and cutoffs $\sigma, \kappa$.
\end{Prop}
{\it Proof.} See next subsection. $\Box$ \medskip

Using  this proposition we infer  the following assertions.
\begin{Thm}\label{LExMassless}
Assume  (V'.1), (V'.2), and (N'). Suppose that the infrared cutoff
 $\sigma=0$ holds.
 If $P\in\Lambda$ and $|P|< \nmass$, then $H(P)$ has a ground state.
\end{Thm}

\begin{Thm}\label{LMainTheorem}Assume  (V'.1), (V'.2), and (N'). Suppose
 that the infrared cutoff $\sigma=0$ holds.  Moreover,
suppose    $E_{\mathrm{bin}}>0$, and  $|P|<\min\big\{\nmass,\sqrt{2\nmass
 E_{\mathrm{bin}}}\big\}$.
Then $H(P)$ has a ground state.
\end{Thm}

\begin{Thm}\label{LIontype}
Assume  (V'.1), (V'.2) and that the system is not neutral in the
 sense that (N') does not hold. Suppose that $\sigma>0$. Then $H(P)$ has a
 ground state for  $P\in\Lambda$ and
 $|P|<m_0$.  Moreover if
 $E_{\mathrm{bin}}>0$, then $H(P)$ has a ground state for 
$|P|<\min\{m_0,\sqrt{2m_0E_{\mathrm{bin}}}\}$.
\end{Thm}

Let $h_N$ be the Hamiltonian $H_N$ ignoring  the quantized radiation field,
i.e.,
\[
 h_N=-\sum_{j=0}^N\frac{\Delta_j}{2m_j}+V.
\]
For $h_N$  one can  define an  binding energy
$\mathsf{e}_{\mathrm{bin}}$ in correspondence to  $E_{\mathrm{bin}}$, see Appendix \ref{Binding}.

\begin{Prop}\label{Observation}For all $\sigma,\kappa$  with
 $0\le\sigma<\kappa<\infty$, one has
\[
E_{\mathrm{bin}}\ge
 \mathsf{e}_{\mathrm{bin}}-\alpha(\kappa^2-\sigma^2)
\]
with  $\alpha=\pi\sum_{j=0}^N(e_j^2/16\pi^2m_j)$.
Thus if $\mathsf{e}_{\mathrm{bin}}>0$ and $\kappa^2-\sigma^2<\mathsf{e}_{\mathrm{bin}}/\alpha$,
then $H(P)$ has a ground state for $|P|<\min\{m_0, \sqrt{2m_0E_{\mathrm{bin}}}\}$.
\end{Prop}
{\it Proof.}\ \ Let $h_{\beta}$ be the Hamiltonian $H_{\beta}$ omitting
the 
quantized radiation field and
$E(h_{\beta})=\inf\mathrm{spec}(h_{\beta})$
 (see Appendix \ref{Binding} for details).
By the diamagnetic inequality (see, e.g. \cite{Hiroshima}), one
concludes
\[
 E_{\beta}\ge E(h_{\beta})
\] 
for all $\beta\in\Pi_N$.
On the other hand, for $f\in\D(-\Delta)$ with $\|f\|=1$,
\[
 E_{N}\le \la f\otimes\Omega, H_{N} f\otimes\Omega\ra=\Big\la f,\Big[-\sum_{j=0}^N\frac{1}{2m_j}\Delta_j+V+\alpha(\kappa^2-\sigma^2)\Big]f\Big\ra,
\] 
which implies
\[
  E_{N}\le E(h_N)+\alpha(\kappa^2-\sigma^2),
\]
where $E(h_N)=\inf \mathrm{spec}(h_N)$.
Combining both  results yields  the assertion.  $\Box$\medskip\\
{\sf Example}
We consider  the hydrogen atom, i.e.,
$N=1$ and $V_{01}(x_0-x_1)=\\-e^2/4\pi|x_0-x_1|$\ $(e_0=-e,\
 e_1=e)$. The system is neutral and  we allow  $\sigma=0$.
By Proposition \ref{Observation},
 one concludes that $E_{\mathrm{bin}}>0$ if 
\begin{align}
 \frac{\mu e^4}{32\pi^2}-\frac{e^2\kappa^2}{16\pi^2\mu}>0\label{RoughEst2}
\end{align}
with $1/\mu=1/\nmass+1/m_1$, because
 $\mathsf{e}_{\mathrm{bin}}=E(h_{\{0\}})+E(h_{\{1\}})-E(h_1)=-E(h_1)=\mu e^4/32\pi^2$.
This rough estimate provides  us with the following imformation.

\begin{itemize}
\item[(1)]
In  case of  hydrogen in nature $e^2/4\pi\simeq 1/137$ and  the ultraviolet cutoff $\kappa$
 must satisfy
\[
 \kappa<\sqrt{\frac{2\pi}{137}}\mu.
\]
\item[(2)]
If we regard $e$ as the coupling parameter, $E_{\mathrm{bin}}>0$
provided
\[
 \frac{\sqrt{2}\kappa}{\mu}<e.
\]
The stronger the coupling $e$, the  larger the  admissible ultraviolet cutoff $\kappa$. 
\end{itemize}

\begin{rem}{\rm
In \cite{FGRS} the binding condition $\mathsf{e}_{\mathrm{bin}}>0$ has
  been proven  for
  the hydrogen molecule $\mathrm{H}_2$ with spin $0$ nuclei. 
The antisymmetry of the electronic part of the wave function can be
  absorbed  into a spin singlet state. Using this result, 
Theorem \ref{LMainTheorem} implies the existence  of the ground state for a
  hydrogen-like molecule coupled to the radiation field, provided
  $\kappa$
is not too large.
Thus if
  $\kappa$ is not too large, also the hydrogen molecule coupled to the
  radiation field  has a ground state.  
}
\end{rem}

\subsection{Proof of Proposition \ref{EInq}}
We will treat the case $m=0$ for the  notational convenience.
All arguments   hold  for  $m>0$, also.
 Let
 $\mathscr{W}=\oplus^{3}L^2(\BbbR^3)$
and $q$ be the bilinear form defined by 
\[
 q(f,g)=\frac{1}{2}\sum_{\mu,\nu=1}^3\int_{\BbbR^3}d_{\mu\nu}(k)\hat{f}_{\mu}(k)\hat{g}_{\nu}(k)\,
 \mathrm{d}k,\ \ f,g\in\mathscr{W},
\]
where $d_{\mu\nu}(k)=\sum_{r=1,2} e_{\mu}^r(k)e_{\nu}^r(k)=\delta_{\mu\nu}-k_{\mu}k_{\nu}/|k|^2$.
Let $(Q,\mu)$ be the probability measure space  for  the mean zero
Gaussian random variables $\{\phi(f)\, |\, f\in\mathscr{W}\}$ with 
covariance  given by 
\[
 \int_Q\phi(f)\phi(g)\, \mathrm{d}\mu(\phi)=\frac{1}{2}q(f,g).
\]
 The photon Fock space  $\mathcal{F}$ can be  naturally  identified
 with $L^2(Q,\mathrm{d}\mu)$ \cite{Hiroshima2}. 
 This representation is called the Schr\"{o}dinger representation.
 Under this identification  
$\Hil^{M}\cong L^2(\BbbR^{3M}\times Q,\mathrm{d}x\otimes\mathrm{d}\mu)$
for arbitrary $M\in\BbbN$.
The unitary operator from $\Hil^M$ to
 $ L^2(\BbbR^{3M}\times Q,\mathrm{d}x\otimes \mathrm{d}\mu)$
 corresponding to  this natural  identification is denoted by $\tilde{S}_M$.

Let $(\mathcal{X},\nu)$ be a $\sigma$-finite measure
space. $f\in L^2(\mathcal{X},\nu)$ 
is called positive if $f$ is nonnegative a.e. and is not the zero
function. A bounded operator $A$ is positivity preserving if
$\la f_1,Af_2\ra\ge 0$ 
for all positive $f_1$ and $f_2\in L^2(\mathcal{X},\mathrm{d}\nu)$.
If $A$ is positivity preserving,
\begin{eqnarray} 
|Af|\le A|f|\ \ \ \mathrm{a.e.}
\end{eqnarray}
for any $f\in L^2(\mathcal{X},\mathrm{d}\nu)$\cite{Gross,Simon1}.
One advantage  of the Schr\"{o}dinger representation is the following
fact: the operator 
$\mathsf{S}_{M+1}\, \ex^{-tH_M}\mathsf{S}_{M+1}^*$ is a  positivity preserving operator in 
$L^2(\BbbR^{3(M+1)}\times Q,\mathrm{d}x\otimes\mathrm{d}\mu)$, where 
$\mathsf{S}_{M+1}=\tilde{S}_{M+1}\, \exp\{\im\frac{\pi}{2}\one\otimes \Num\}$
\cite{Hiroshima1}.

From now on, we  fix $N\in\BbbN$ arbitrarly and   denote
 $\mathsf{S}=\mathsf{S}_{N}$ for  notational simplicity.

\begin{lemm}\label{PP1}
Let $\mathscr{V}(P)$ and $K(P)$ be the operators defined by (\ref{VofP}) and
 (\ref{KofP}) respectively.
\begin{itemize}
 \item[{\rm(i)}]   $\mathsf{S}\mathscr{V}(0)\mathsf{S}^*$ is positivity
 preserving.
\item[{\rm(ii)}]   $\mathsf{S}\, \ex^{-sK(0)}\mathsf{S}^* $ is positivity
		   preserving for all $s>0$.
\end{itemize}
\end{lemm}
$Proof.$\ \ (i)\  Since
$\exp\{\im x_1\cdot \nabla_j\otimes\one\}$ and
$\exp\{\im x_1\cdot \one\otimes P_{\mathrm{f}}\}$
are  translations, the result follows.

(ii)\ Note that
 $\mathsf{S}\mathscr{V}(0)\mathsf{S}^*,\mathsf{S}\mathscr{V}(0)^*\mathsf{S}^*$
 and $\mathsf{S}\, \ex^{-sH_A}\mathsf{S}^*$ are
  positivity preserving. Thus $\mathsf{S}\, \ex^{-s K(0)}\mathsf{S}^*$
 is also  positivity preserving by the fact
 $\ex^{-sK(0)}=\mathscr{V}(0)\,\ex^{-s H_A}\mathscr{V}(0)^*$. $\Box$

\begin{lemm}\label{PP2} For all $P\in\BbbR^3$, the following holds.
\begin{itemize}
\item[{\rm (i)}]   $|\mathsf{S}\, \mathscr{V}(P)\mathsf{S}^* F| \le \mathsf{S}\, \mathscr{V}(0)\mathsf{S}^* |F|$ a.e..
\item[{\rm (ii)}]  $|\mathsf{S}\, \ex^{-sK(P)}\mathsf{S}^* F| \le \mathsf{S}\, \ex^{-sK(0)}\mathsf{S}^* |F|$ a.e..
\end{itemize}
\end{lemm}
$Proof.$\ \ (i)\ \  For a.e. $x$ and $\phi$,
\begin{align*}
|(\mathsf{S}\, \mathscr{V}(P)\mathsf{S}^* F)(x,\phi)| &=|\ex^{\im x_1\cdot
 P}(\mathsf{S}\, \mathscr{V}(0)\mathsf{S}^* F)(x,\phi)|\\
&\le |(\mathsf{S}\, \mathscr{V}(0)\mathsf{S}^* F)(x,\phi)|\\
&\le (\mathsf{S}\, \mathscr{V}(0)\mathsf{S}^* |F|)(x,\phi)
\end{align*}
 by Lemma \ref{PP1}.

(ii) By (i), Lemma \ref{PP1}, and the fact that
$\mathsf{S}\,\ex^{-sH_A}\mathsf{S}^*$ 
is positivity preserving,
\begin{align*}
|\mathsf{S}\, \ex^{-sK(P)}\mathsf{S}^* F|&=|\mathsf{S}\,
 \mathscr{V}(P)\mathsf{S}^*\mathsf{S}\,
 \ex^{-sH_A}\mathsf{S}^*\mathsf{S}\, \mathscr{V}(P)^*\mathsf{S}^* F|\\
&\le \mathsf{S}\,
 \mathscr{V}(0)\mathsf{S}^*|\mathsf{S}\,
 \ex^{-sH_A}\mathsf{S}^*\mathsf{S}\, \mathscr{V}(P)^*\mathsf{S}^* F|\\
&\le  (\mathsf{S}\mathscr{V}(0)\mathsf{S}^*)(\mathsf{S}
 \ex^{-sH_A}\mathsf{S}^*)|\mathsf{S}\mathscr{V}(P)^*\mathsf{S}^* F|\\
&\le  (\mathsf{S}\mathscr{V}(0)\mathsf{S}^*)(\mathsf{S}
 \ex^{-sH_A}\mathsf{S}^*)(\mathsf{S}\mathscr{V}(0)^*\mathsf{S}^*)| F|\\
&=\mathsf{S}\, \ex^{-sK(0)}\mathsf{S}^* |F|
\end{align*}
for a.e.. $\Box$

\begin{Prop}\label{Dimag}
For all $t>0$ and $P\in\BbbR^3$,
\[
 |\mathsf{S}\, \ex^{-tH(P)}\mathsf{S}^*F|\le \mathsf{S}\,
  \ex^{-tH(0)}\mathsf{S}^*|F|\ \ \mbox{a.e..}
\]
\end{Prop}
$Proof.$\ \ Let $A_n(P)=(\ex^{-tH_{\mathrm{PF}}/n}\, \ex^{-tK(P)/n})^n$
for all $n\in\BbbN$. By Kato's strong product formula \cite[Theorem
S.21]{ReSi1}, $\slim A_n(P)=\ex^{-t H(P)}$. For all $n\in\BbbN$,
\[
 |\mathsf{S}\, A_n(P)\mathsf{S}^*F|\le \mathsf{S}\, A_n(0)\mathsf{S}^*|F|
\]
by Lemma \ref{PP2} and the fact that
$\mathsf{S}\,\ex^{-sH_{\mathrm{PF}}}\mathsf{S}^*$ 
is positivity preserving. Taking the limit $n\to \infty$, we get the
desired result. $\Box$ 
\medskip\\
{\sf Proof of Proposition \ref{IneqGSE1}}.\\
By Proposition \ref{Dimag} we get
\[
 \la F, \mathsf{S}\, \ex^{-tH(P)}\mathsf{S}^* F \ra\le \la |F|,
  \mathsf{S}\, \ex^{-t H(0)}\mathsf{S}^*|F|\ra.
\]
for $F\in  L^2(\BbbR^{3N}\times Q,\mathrm{d}x\otimes\mathrm{d}\mu)$.
From this we immediately obtain the desired result. $\Box$

\appendix

\section{Proof of Proposition \ref{Loss}}
We start  with a  lemma about convex functions.
Denote by  $\mathcal{C}$ the set of convex functions $g:\BbbR^n\to\BbbR$
that satisfy $0\le g(x)\le |x|^2/2$.

\begin{lemm}\label{convex} 
Fix any two points $P$ and $Q$ in $\BbbR^n$. Then the function
\[
 \Delta(P,Q):=\sup\{g(P)-g(Q)\, :\, g\in\mathcal{C}\}
\]
equals
\begin{align*}
\Delta(P,Q)= \begin{cases}
Q\cdot(P-Q)+|P-Q||Q|, &\mbox{if $|P-Q|\le |Q|$,}\\
P^2/2, & \mbox{if $|P-Q|\ge |Q|$. }
\end{cases}
\end{align*}
Moreover  the maximizer  is given by
\begin{align*}
g(x)= \begin{cases}
Q\cdot (x-Q)+|x-Q||Q|, &\mbox{if $|x-Q|\le |Q|$,}\\
|x|^2/2, & \mbox{if $|x-Q|\ge |Q|$. }
\end{cases}
\end{align*}
\end{lemm}
{\it Proof.}\ \ First we set $g(Q)=A$ where $A>0$ is an arbitrary number
less than $Q^2/2$. Next we consider all the rays starting  at $(Q,A)$
that are tangent to the surface $z=x^2/2\ (x\in\BbbR^n)$. Such a ray is
given  in parametrized form by
\[
 x(t)=Q+te,\ \ z(t)=A+tE,
\]
where $e$ is a unit vector in $\BbbR^n$ and $E$ is a real number. As we
said, this ray has to touch the surface at the point  $(Q+t_0e, A+t_0E)$
which means that 
\[
 A+t_0E=(Q+t_0e)^2/2
\]
together with the tangency condition $(e,E)\bot (Q+t_0e, -1)$.
From this one sees  that 
\[
 t_0^2=Q^2-2A>0
\]
and 
\[
 E=Q\cdot e +t_0.
\]
Thus, for every direction $e$ there are two touching points
\[
x_0=Q\pm e\sqrt{Q^2-2A},\ \ z_0=Q^2-A\pm Q\cdot e\sqrt{Q^2-2A}. 
\]
Note that the $x$ components of the touching points sit on a sphere in
$\BbbR^n$ given by the equation $(x-Q)^2=Q^2-2A$.

The point about these touching segments is the following. Every
function $g\in\mathcal{C}$ with $g(Q)=A$ must have its graph below this
segment, in other words
\[
 g(Q+te)\le A+tE
\]
for all $t$ with $t^2\le Q^2-2A$. Thus, if $P$ is inside the sphere,
i.e., 
\[
 (P-Q)^2\le Q^2-2A,
\]
we have that $P=Q+te$ and hence
\[
 g(P)\le A+tE=A+t(Q\cdot e+t_0)=A+Q\cdot (P-Q)+|P-Q|\sqrt{Q^2-2A},
\]
noting that $t$ and $t_0$ need to have the same sign. Thus 
\[
 g(P)-g(Q)\le Q\cdot (P-Q)+|P-Q|\sqrt{Q^2-2A}.
\]

Next we consider the case $P$ is outside the sphere. Clearly in this
case  the largest
value for $g(P)$ is  $P^2/2$ and hence in this case
\[
 g(P)-g(Q)\le P^2/2-A.
\]
Thus we have that
\begin{align*}
g(P)-g(Q)\le  \begin{cases}
Q\cdot (P-Q)+|P-Q|\sqrt{Q^2-2A}, &\mbox{if $(P-Q)^2\le Q^2-2A$,}\\
P^2/2-A, & \mbox{if $(P-Q)^2\ge Q^2-2A$. }
\end{cases}
\end{align*}
Note that for $(P-Q)^2=Q^2-2A$ we find that
\[
 Q\cdot (P-Q)+|P-Q|\sqrt{Q^2-2A}=P^2/2-A.
\]

Next we claim that
\begin{align*}
g(P)-g(Q)\le \begin{cases}
Q\cdot(P-Q)+|P-Q||Q|, &\mbox{if $(P-Q)^2\le Q^2$,}\\
P^2/2, & \mbox{if $(P-Q)^2\ge Q^2$. }
\end{cases}
\end{align*}
This is obvious on the set of all $Q$s with $(P-Q)^2 \le Q^2-2A$ and
for all those that satisfy $(p-Q)^2 \ge Q^2$.
Thus, it remains to show that for all $Q$s that satisfy $Q^2-2A\le (P-Q)^2\le Q^2$,
\[
 P^2/2-A\le Q\cdot (P-Q)+|P-Q||Q|,
\]
which is the same as 
\[
 (|Q|-|P-Q|)^2/2\le A.
\]
Since $|P-Q|\le |Q|$ it suffices to show that
\[
 |Q|-|P-Q|\le \sqrt{2A}
\]
or 
\[
 |P-Q|\ge |Q|-\sqrt{2A}.
\]
Since, by assumption $|P-Q|\ge \sqrt{Q^2-2A}$ this follows once we show
that 
\[
 \sqrt{Q^2-2A}\ge |Q|-\sqrt{2A}.
\]
Squaring both sides yields
\[
 Q^2-2A\ge Q^2-2|Q|\sqrt{2A}+2A
\]
or equivalently
\[
 |Q|\ge \sqrt{2A}
\]
which follows from the fact that $A\le Q^2/2$. $\Box$\medskip\\
{\sf Proof of Proposition  \ref{Loss}}

Write $F(P)$ as 
\[
 F(P)=\frac{P^2}{2}+F(0)-h(P)
\]
where $h(P)$ is convex. From (b) in Proposition \ref{Loss} we get $h(P)\ge
0$ and from (a) we learn  that $h(P)\le P^2/2$. Hence
\begin{align*}
 F(P-k)-F(P)&=\frac{(P-k)^2}{2}-\frac{P^2}{2}-[h(P-k)-h(P)]\\
&=-k\cdot P+\frac{k^2}{2}-[h(P-k)-h(P)].
\end{align*}
Using the lemma above we get
\begin{align*}
h(P-k)-h(P)\le \begin{cases}
-P\cdot k +|k||P|, &\mbox{if $|k|\le |P|$,}\\
(P-k)^2/2, & \mbox{if $|k|\ge |P|$ }
\end{cases}
\end{align*}
and hence
\begin{align*}
F(P-k)-F(P)&\ge -k\cdot P+\frac{k^2}{2}-\begin{cases}
-P\cdot k +|k||P|, &\mbox{if $|k|\le |P|$,}\\
(P-k)^2/2, & \mbox{if $|k|\ge |P|$ }
\end{cases}\\
&=\begin{cases}
-|k||P|+\frac{k^2}{2}, &\mbox{if $|k|\le |P|$,}\\
-\frac{P^2}{2}, & \mbox{if $|k|\ge |P|$. }
\end{cases}
\end{align*}
This proves the proposition. $\Box$

\section{Proof of Proposition \ref{PFBinding}}\label{PrBC}

 In order to clarify the dependence of $\Nm$,
we denote our Hamiltonian by $H(P;\Nm)$ instead of $H(P)$.
Also we denote the bottom of spectrum of  $H(P; \Nm)$ by $E(P;\Nm)$.

\begin{lemm}\label{static}
\begin{itemize}
\item[{\rm (i)}] $E(P;\Nm)\to E_{ N}^{\infty}$ as $\Nm\to \infty$.
\item[{\rm (ii)}] $\Sigma(P;\Nm)\ge \Sigma_{ N}^{\infty}$ for all
		 $\Nm$ and $P$, where $\Sigma(P;\Nm)$ and $\Sigma_{N}^{\infty}$
 denote the threshold energy correspoinding to  $H(P; \Nm)$ and 
$H_{ N}^{\infty}$ which are similarly defined by (\ref{Threshold}).
\end{itemize}
\end{lemm}
{\it Proof.}\ \ (i) For all $\Nm>0$, $H(P;\Nm)$ and $H_{ N}^{\infty}$
are both essentially self-adjoint on $\Hil_{\mathrm{fin}}^N$ by
Proposition \ref{SAPF} and Theorem \ref{SA}. Moreover,
for all  $\vphi\in\Hil_{\mathrm{fin}}^N$, $H(P;\Nm)\vphi\\ \to
H_{N}^{\infty}\vphi$
 as  $\Nm\to \infty$. Therefore $H(P;\Nm)\to H_{N}^{\infty}$
in the strongly resolvent sense by \cite[Theorem VIII.25]{ReSi1} which
implies the  desired result by \cite[Theorem VIII.24]{ReSi1}.

(ii) This follows from the operator inequality $H(P;\Nm)\ge
H_{N}^{\infty}$. $\Box$\medskip\\
{\sf Proof of Proposition \ref{PFBinding}}\\
 Note first that, by \cite{Grie}, 
\[
 \Sigma_{ N}^{\infty}=\min\big\{E_{\beta}^{\infty}+E_{\bar{\beta}}^{\infty}\, |\,
 \beta\subset \{1,\dots, N\}\, \mathrm{and}\, \beta\neq\emptyset,
 \{1,\dots,N\}\big\}.
\]
By Lemma \ref{static} (i), there is a $\Nm>0$ such that 
\[
 E(0;\Nm)-E_{N}^{\infty}<\frac{E_{\mathrm{bin}}^{\infty}}{2}.
\]
(Remark, here, that  $E(0;\Nm)\ge E_{N}^{\infty}$.)
Hence, by Lemma \ref{static} (ii) and Theorem \ref{IneqGSE1} (ii),
\begin{align*}
\Sigma(P;\Nm)-E(P;\Nm)  &\ge
 \Sigma_{N}^{\infty}-E(0;\Nm)-\frac{P^2}{2\Nm}\\
 &\ge\Sigma_{N}^{\infty}-E_{N}^{\infty}-\frac{P^2}{2\Nm}-\frac{E_{\mathrm{bin}}^{\infty}}{2} \\
&= \frac{E_{\mathrm{bin}}^{\infty}}{2}-\frac{P^2}{2\Nm}. 
\end{align*}
Thus if $|P|<\sqrt{\Nm E_{\mathrm{bin}}^{\infty}}$, the binding condition
(B.C.) follows. $\Box$

\section{A uniform   estimate for $\mathfrak{P}_j$}

Let
\begin{align*}
\mathfrak{P}_j&=-\im\nabla_j\otimes\one+e A(x_j),\ \ j=1,\dots,N,\\
\mathfrak{P}_0&=P+\im\sum_{j=1}^{N}\nabla_j\otimes\one-\one\otimes
 P_{\mathrm{f}}-Ze A(0).
\end{align*}

\begin{lemm}\label{UniEst}
For each $j=0,1,\dots,N,\ l=1,2,3$ and $\vphi\in\D(H_m(P))$, there is a constant $C$ independent of $m$ and
 $P$ such  that 
\[
 \|\mathfrak{P}_{j,l}\vphi\|\le C\big(\|H_{\Mass}(P)\vphi\|+\|\vphi\|\big).
\]
\end{lemm}
{\it Proof.}\ \
Throughout  this proof, we use the symbol  $H^{V=0}(P)$ which
 means the Hamiltonian (\ref{Hami}) with  $V=0$. 
By Lemma \ref{IndPIneq} (and the fact $\tilde{H}(P)=H(P)$ and $\tilde{H}^{V=0}(P)=H^{V=0}(P)$), there is
a constant $C_1>0$ independent of $m$ and $P$ so that 
\[
 \la \vphi, H^{V=0}(P)\vphi\ra\le C_1\big(\la\vphi,H(P)\vphi\ra+\|\vphi\|^2\big)
\]
for  $\vphi\in\Hil^{N}_{\mathrm{fin}}$.  
 Since $H(P)\le H_{\Mass}(P)$, we have
\begin{align*}
 \|\mathfrak{P}_{j,l}\vphi\|^2&\le \la \vphi, H^{V=0}(P)\vphi\ra\\
&\le C_1\big(\la \vphi,H(P)\vphi\ra+\|\vphi\|^2\big)\\
&\le C_1\big(\la \vphi,H_{\Mass}(P)\vphi\ra+\|\vphi\|^2\big)\\
&\le 2C_1 \big(\|H_{\Mass}(P)\vphi\|^2+\|\vphi\|^2\big).
\end{align*}
Since $\Hil_{\mathrm{fin}}^N$ is a core of $H_m(P)$, the lemma follows. $\Box$

\section{Binding condition  for the  Schr\"odinger atom}\label{Binding}
We consider the $(N+1)$-particle Schr\"odinger operator acting in
$L^2(\BbbR^{3(N+1)})$  given by
(\ref{PHamiltonian}), to repeat,
\begin{align}
 h_N=-\sum_{j=0}^{N}\frac{1}{2m_j}\Delta_j+\sum_{0\le i <j\le N}V_{ij}(x_i-x_j),\label{AtomHami}
\end{align}
where $m_0=m_{\mathrm{n}}$ and $ m_j=\emass$ for
$j=1,\dots,N$.
The purpose of this appendix is to prove that in this case our binding
condition (B.C.) reduces to the conventional binding condition.

Let $R$ be the center of mass
\[
 R=\frac{1}{\tmass}\sum_{j=0}^Nm_jx_j
\]
and define $\mathcal{X}=\{x\in\BbbR^{3(N+1)}\, |\,
R=0\}$.
In the $3N$-dimensional vector space $\mathcal{X}$,
we use  atomic coordinates $y_i=x_i-x_0,\ i=1,\dots,N$.
Since we can identify $\mathcal{X}$ as $\BbbR^{3N}$ under  atomic coordinates, we obtain the
following identification
\[
 L^2(\BbbR^{3(N+1)})=L^2(\mathcal{X}^c)\otimes L^2(\mathcal{X})=L^2(\BbbR^3)\otimes L^2(\BbbR^{3N}).
\]
Moreover, our Hamiltonian can be  expressed as 
\begin{align}
h_N=-\frac{1}{2\tmass}\Delta_R\otimes\one+\one\otimes \tilde{h},\label{NB}
\end{align}
where
\begin{align}
\tilde{h}&=-\sum_{j=1}^N\frac{1}{2\mu_j}\Delta_{y_j}+\sum_{i<j}\frac{1}{m_0}\nabla_{y_i}\cdot\nabla_{y_j}+\tilde{V},\
 \ \frac{1}{\mu_j}:=\frac{1}{m_0}+\frac{1}{m_j},\label{CenterRM}
 \\
\tilde{V}(y_1,\dots,y_N)&=\sum_{j=1}^NV_{0j}(y_j)+\sum_{1\le i<j\le N}V_{ij}(y_i-y_j).\nonumber
\end{align}
Remark that the total momentum $\Pt=\sum_{j=0}^N(-\im\nabla_j)$ is
represented by $\Pt=-\im\nabla_R$ in our coordinates.
Let $\mathcal{F}$ be the Fourier transformation with respect to $R$.
 Clearly $\mathcal{F}$ is unitary and
$\mathcal{F}\Pt\mathcal{F}^*=k$ (as multiplication operator). 
Thus $\mathcal{F}$ yields  a spectral representaiton of
$\Pt$. Furtheremore, by (\ref{NB}), one obtains
\[
\mathcal{F}h_N\mathcal{F}^*=\frac{k^2}{2\tmass}\otimes\one+\one\otimes \tilde{h}. 
\] 
Thus we are lead to  the following fibre direct integral represetation of
$\mathcal{F}h_N\mathcal{F}^*$,
\begin{align*}
 \mathcal{F}h_N\mathcal{F}^*=\int^{\oplus}_{\BbbR^3}h(P)\,
 \mathrm{d}P,\ \ 
h(P)=\frac{P^2}{2\tmass}+\tilde{h}.
\end{align*}

For a self-adjoint operator $A$ on $L^2(\BbbR^{3d})$, we introduce
\[
 \Sigma(A)=\lim_{R\to\infty}\Big(\inf_{\vphi\in\mathcal{D}_{d,R},\, \|\vphi\|=1}\la\vphi,A\vphi\ra\Big),
\]
where $\mathcal{D}_{d,R}=\{\vphi\in C^{\infty}_0(\BbbR^{3d})\, |\,
\vphi(x)=0,\, \mathrm{if}\, |x|<R\}$.
For a self-adjoint  operator $A$ bounded
from below, $E(A)$ stands for $\inf \mathrm{spec}(A)$.

\begin{Prop}\label{BinBin}
For all $P$,
\begin{align*}
 \Sigma(h(P))-E(h(P))&=\inf\mathrm{ess.\, spec}(h(P))-E(h(P))\\
&=\inf\mathrm{ess.\, spec}(\tilde{h})-E(\tilde{h}),
\end{align*}
where $\mathrm{ess.\,  spec}(A)$ means the essential spectrum of the linear
 operator $A$.
Thus, if $\Sigma(h(P))-E(h(P))>0$ for some $P$, then $h(P)$ has a ground
 state for all $P$.
\end{Prop}
{\it Proof.}\ \ 
Clearly
\[
 \Sigma(h(P))=\frac{P^2}{2\tmass}+\Sigma(\tilde{h}),\ \ \ E(h(P))=\frac{P^2}{2\tmass}+E(\tilde{h}).
\]
We also note that 
$\Sigma(\tilde{h})=\inf \mathrm{ess.\,  spec}(\tilde{h})$
by \cite{Persson}.
Hence
\begin{align*}
\Sigma(h(P))-E(h(P))&=\inf\mathrm{ess.\,  spec}(\tilde{h})-E(\tilde{h})\\
&=\inf\mathrm{ess.\,  spec}(h(P))-E(h(P)).\ \ \Box
\end{align*}

Let $\Pi_N$ be the set of the subsets of $\{0,1,2,\dots,N\}$. We denote by
$h_{\beta}$ the Hamiltonian of the form (\ref{AtomHami}), but only for the
particles in the set $\beta$,
\[
 h_{\beta}=-\sum_{j\in\beta}\frac{1}{2m_j}\Delta_j+\sum_{i,j\in\beta,0\le i<j\le N}V_{ij}.
\]
The binding energy $\mathsf{e}_{\mathrm{bin}}$ for $h_N$ is  defined by
\[
 \mathsf{e}_{\mathrm{bin}}=\min\{E(h_{\beta})+E(h_{\bar{\beta}})\, |
 \, \beta\in\Pi_N,\ \beta\neq \emptyset, \{0,1,\dots,N\}\}-E(h_N)
\]

\begin{Prop}
For all $P$,
\[
 \mathsf{e}_{\mathrm{bin}}= \Sigma(h(P))-E(h(P)).
\]
Thus, if $\mathsf{e}_{\mathrm{bin}}>0$, $h(P)$ has a ground state for all $P$. 
\end{Prop}
{\it Proof.}\ \ 
By the HVZ-theorem \cite[Theorem 3.7]{CFKS}, we have
\[
  \mathsf{e}_{\mathrm{bin}}=\inf\mathrm{ess.\, spec}(\tilde{h})-E(\tilde{h})
\]
and the assertion follows from Proposition \ref{BinBin}.
$\Box$


\begin{thebibliography}{100}
\bibitem{AGG1}         L. Amour, B. Grebert, J. Guillot,  The
	dressed  nonrelativistic electron in a magnetic field,
	C. R. Math. Acad. Sci. Paris 340 (2005) 421-426.
\bibitem{AGG}         L. Amour, B. Grebert, J. Guillot,  The
	dressed mobile atoms and ions, preprint arXiv:math-ph/0507052.
\bibitem{AMZ}         N. Angelescu, R. A. Minlos, V. A. Zagrebnov,
	  Lower spectral branches of a particle coupled to a bose
	field, Rev. Math. Phys. 13 (2005) 1111-1142.
\bibitem{AH1}         A. Arai, M. Hirokawa,  On the existence 
  and uniqueness of ground states of a generalized spin-boson model,
J.  Funct. Anal.   151 (1997) 455-503.
\bibitem{ARV}         E. A. G. Armour, J.-M. Richard, K. Varga,
	 Stability of few-charged system in quantum mechanics,
Phys.  Rep.   413  (2005)  1-90.
\bibitem{Chen}      V. Bach, T. Chen, J. Fr\"ohlich,  I. M. Sigal,
          The renormalized electron mass in non-relativisitc quantum
        electrodynamics,
         preprint arXiv:math-ph/0507043.
\bibitem{BFS}      V. Bach, J. Fr\"ohlich, I. M. Sigal,  Quantum
	electrodynamics of confined non-relativistic particles,
	Adv. Math.  137 (1998) 299-395.
\bibitem{BFS2}     V. Bach, J. Fr\"ohlich, I. M. Sigal,  Spectral
	analysis for systems of atoms and molecules coupled to the 
        quantized radiation field, Commun. Math. Phys.  207 (1999)
        249-290.
\bibitem{CFKS}        H. L. Cycon, R. G. Froese, W. Kirsch, B. Simon,
	Schr\"odinger Operators, Springer-Verlag (1987).
\bibitem{DG1}         J. Derezi\'{n}ski, C. G\'erard,
          Asymptotic completeness in quantum field theory.
        Massive Pauli-Fierz Hamiltonians,
         Rev. Math. Phys.  11 (1999) 383-450.
\bibitem{Froehlich}   J. Fr\"ohlich,  Existence of dressed one
	electron states in a class of persistent models, Fortschritte
	der Physik   22 (1974) 159-198.
\bibitem{FGS}   J. Fr\"ohlich,  M. Griesemer,  B. Schlein, 
	Asymptotic completeness for Compton scattering,
	Comm. Math. Phys.  252 (2004) 415-476.
\bibitem{FGS2}   J. Fr\"ohlich, M. Griesemer, B. Schlein, 
	Rayleigh scattering at atoms with dynamical nuclei, preprint
	arXiv:math-ph/0509009.
\bibitem{FGRS}   J. Fr\"ohlich, G.-M. Graf, J.-M. Richard, 
	M. Seifert,  Proof of stability of the hydrogen molecule,
	Phys. Rev. Lett.  71 (1993) 1332-1334.
\bibitem{Ge1} C. G\'erard,  On the existence of 
ground states for massless Pauli-Fierz Hamiltonians, Ann. Henri
 Poincar\'e   1 (2000) 443-458.
\bibitem{Grie}  M. Griesemer,  Exponential decay and ionization
	thresholds in non-relativistic quantum electrodynamics,
	J. Funct.  Anal.
	 210 (2004)  321-340.
\bibitem{LLG}         M. Griesemer, E. H. Lieb,  M. Loss,  Ground
	states in non-relativistic quantum electrodynamics,
	Invent. Math. 145 (2001) 557-595.
\bibitem{Gross}    L. Gross,  Existence and uniqueness of physical
	ground states, J. Funct. Anal.  10 (1972) 52-109.
\bibitem{HHS}   M. Hirokawa, F. Hiroshima,  H. Spohn , 
        Ground state for point particles interacting through a massless
	scalar bose field, Adv. in Math.  191 (2005) 339-392. 
\bibitem{Hiroshima}   F. Hiroshima,  Functional integral
       representation of a model in quantum electrodynamics, 
       Rev. Math. Phys.  9 (1997) 489-530.
\bibitem{Hiroshima1}  F. Hiroshima,  Ground states of a model in
	nonrelativistic quantum electrodynamics. II,
	J. Math. Phys. 41 (2000) 661-674.
\bibitem{Hiroshima3}  F. Hiroshima,  Essential self-adjointness of
	translation-invariant quantum field models for arbitrary
	coupling constants, Commun. Math. Phys. 211 (2000) 585-613.
\bibitem{Hiroshima2}  F. Hiroshima,  Self-adjointness of the
	Pauli-Fierz Hamiltonian
       for arbitrary values of coupling constants, Ann. Henri Poincare
	3 (2002) 171-201.
\bibitem{HunSig}     W. Hunziker, I. M. Sigal,  The quantum
 N-body problem,
 J. Math. Phys. 41 (2000) 3448-3510.
\bibitem{LL2}	        E. H. Lieb,  M. Loss, Analysis,
	Graduate Studies in Mathematics, American Mathematical Society,
	1997.
\bibitem{LL}          E. H. Lieb and M. Loss,  Existence of atoms
	and molecules in non-relativistic quantum electrodynamics,
	Adv. Theor. Math. Phys.  7 (2003)  667-710.
\bibitem{Moller}      J. S. M\o ller, The translation invariant massive
	Nelson model I. The bottom of the spectrum, Ann. Henri Poincar\'{e}
	6 (2005) 1091-1135.
\bibitem{Persson}     A. Persson,  Bounds for the discrete part of
	the spectrum of a semi-bounded Schr\"odinger operator,
	Math. Scand.  8 (1960) 143-153.
\bibitem{Pizzo}       A. Pizzo,  One particle (improper) states in
	Nelson's  massless model, Ann. Henri   Poincare 4 (2003)
        439-486.
\bibitem{ReSi1}	M.  Reed and B.  Simon, Methods of Modern
	Mathematical Physics Vol. I,  Academic Press, New York, 1975. 
\bibitem{ReSi2}	M.  Reed and B.  Simon, Methods of Modern
	Mathematical Physics Vol. II,  Academic Press, New York, 1975.
\bibitem{ReSi4}	M.  Reed and B.  Simon, Methods of Modern
	Mathematical Physics Vol. IV, Academic Press, New York, 1978.
\bibitem{Simon}	 B.  Simon, The $P(\phi)_2$ Euclidean
	(Quantum) Field Theory,  Princeton University Press,
	 1974.  
\bibitem{Spohn1}      H. Spohn,  The polaron at large total
	momentum, J. Phys. A  21 (1988) 1199-1211.
\bibitem{Spohn2}      H. Spohn, Dynamics of Charged Particles and
	Their Radiation Field,  Cambridge University Press,  2004. 
\bibitem{Zhislin}      G. Zhislin,  A study of the spectrum of the
	Schr\"odinger operator for a system of several particles,
	Trudy Moscov Mat. Obsc.  9 (1960) 81-120.
\end{thebibliography}
\end{document}